\documentclass[a4paper, 12pt]{article}
\usepackage{fullpage}
\usepackage{siunitx}
\usepackage{listings}
\usepackage{amsmath, bm, amssymb}
\usepackage{graphicx}
\usepackage{setspace}
\usepackage{physics, braket}
\graphicspath{{./pics/}}
\usepackage{cite}
\usepackage[font=it]{caption}
\usepackage{subcaption}
\usepackage{secdot}
\usepackage{hyperref}
\usepackage{authblk}

\numberwithin{equation}{section}

\newcommand{\ud}{\textup{d}}
\newcommand{\ui}{\textup{i}}
\newcommand{\ue}{\textup{e}}
\newcommand{\uh}{\textup{h}}
\newcommand{\uc}{\textup{c}}
\newcommand{\ua}{\textup{a}}
\newcommand{\ub}{\textup{b}}
\newcommand{\uB}{\textup{B}}

\newcommand{\uin}{\textup{in}}
\newcommand{\uout}{\textup{out}}

\title{Autonomous Quantum Heat Engine Based on Non-Markovian Dynamics of an Optomechanical Hamiltonian}

\author[1,*]{\textbf{Miika Rasola}}
\author[1,2,$\dagger$]{\textbf{Mikko Möttönen}}

\affil[1]{QCD Labs, QTF Centre of Excellence, Department of Applied Physics, Aalto University, P.O. Box 13500, FI-00076 Aalto, Finland}
\affil[2]{QTF Centre of Excellence, VTT Technical Research Centre of Finland Ltd., P.O. Box 1000, 02044 VTT, Finland}
\affil[ ]{}
\affil[*]{\textup{miika.rasola@aalto.fi}}
\affil[$\dagger$]{\textup{mikko.mottonen@aalto.fi}}

\begin{document}

\maketitle
\onehalfspacing

\begin{abstract}
We propose a recipe for demonstrating an autonomous quantum heat engine where the working fluid consists of a harmonic oscillator, the frequency of which is tuned by a driving mode. The working fluid is coupled two heat reservoirs each exhibiting a peaked power spectrum, a hot reservoir peaked at a higher frequency than the cold reservoir. Provided that the driving mode is initialized in a coherent state with a high enough amplitude and the parameters of the utilized optomechanical Hamiltonian and the reservoirs are appropriate, the driving mode induces an approximate Otto cycle for the working fluid and consequently its oscillation amplitude begins to increase in time. We build both an analytical and a non-Markovian quasiclassical model for this quantum heat engine and show that reasonably powerful coherent fields can be generated as the output of the quantum heat engine. This general theoretical proposal heralds the in-depth studies of quantum heat engines in the non-Markovian regime. Further, it paves the way for specific physical realizations, such as those in optomechanical systems, and for the subsequent experimental realization of an autonomous quantum heat engine.
\end{abstract}

\section*{Introduction}

The global societal impact of heat engines has been enormous through history. Various types of heat engines such as those powering pumps and motors have been extensively analyzed and utilized in a great number of practical applications since the industrial revolution. The idea of modelling a quantum system as a heat engine was presented already in the 50's, when Scovil and Schulz-DuBois analyzed the three-level maser as a heat engine~\cite{Scovil}. During the recent decades however, first the development of quantum thermodynamics~\cite{qthermo, Goold, gemmer}, and later the rapidly increasing number of quantum technological devices~\cite{Wolfgang}, has led to renewed interest in heat engines operating at the mesoscopic quantum level---quantum heat engines (QHEs).

Recently, QHE behaviour has been experimentally demonstrated in various microscopic atomic devices, such as in single trapped ions~\cite{rossnagel}, a spin coupled to single-ion motion~\cite{vonL, vonH}, nitrogen vacancy center interacting with a light field~\cite{Klatzow}, and in nuclear magnetic resonance~\cite{deAssis, Peterson}. In addition, a QHE driven by atomic collisions was reported in~\cite{Bouton} where a quantum Otto cycle was achieved in large quasi-spin states of Cesium impurities immersed in an ultracold Rubidium bath. Also superconducting circuits provide prospects for demonstrating QHEs~\cite{Pekola07, Pekola2015, Karimi, Thomas, Ronzani, Tan2017}. All of these devices are inherently not autonomous in the sense that they are all driven by some type of external control. In this case, it is very difficult to extract, or even directly observe, the work produced by the heat engine. If we wish to demonstrate an actual autonomous quantum heat engine, efforts should be directed at creating a device, work output of which can be directly observed instead of being superimposed on macroscopic external control fields. In this paper, we propose and theoretically study a device potentially remedying this issue, a topic studied previously in references~\cite{Naseem2019, Izadyari2022, Hardal2017, Mari2015, Gelbwaser-Klimovsky2015, Gelbwaser-Klimovsky_2013} as discussed below.

\begin{figure}[!ht]
\centering
\includegraphics[width=0.6\textwidth, trim={70 390 50 23}, clip]{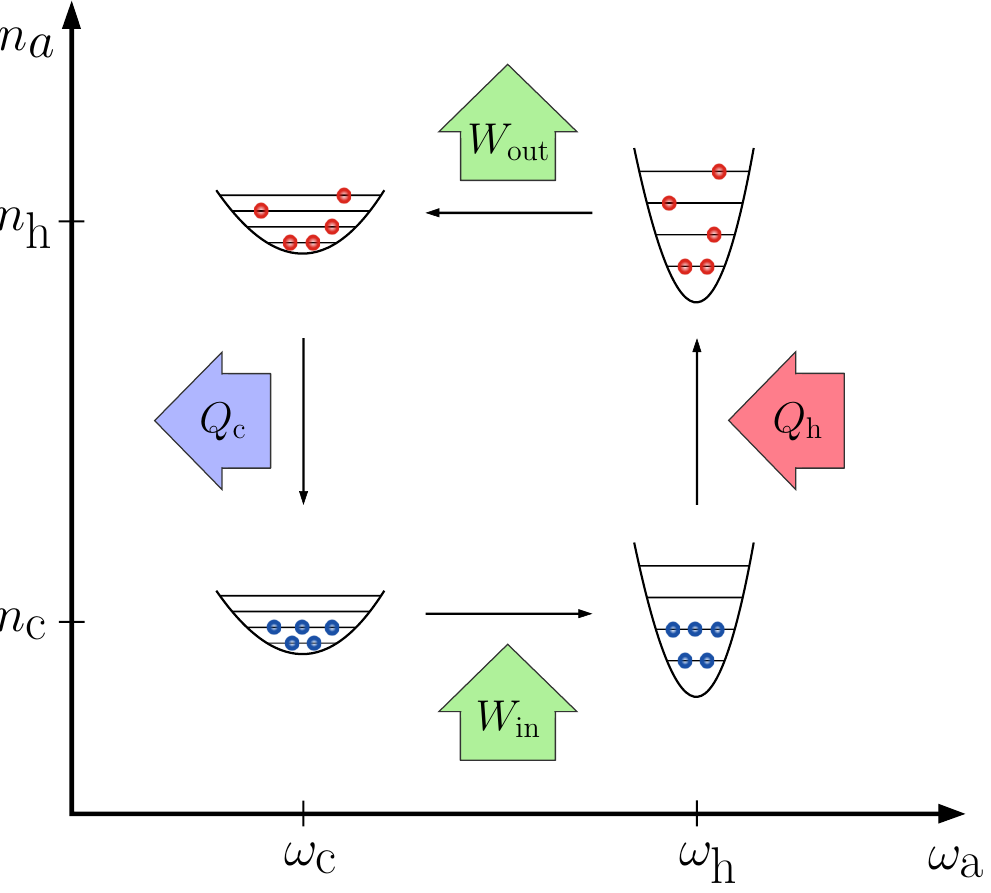}
\caption{\textbf{Quantum Otto cycle considered in this work.} The angular frequency $\omega_\ua$ of the harmonic oscillator depicted as the quadratic potential traverses between values $\omega_\uc$ and $\omega_\uh$ at average photon numbers $n_\uc$ and $n_\uh$, respectively. The work obtained from the cycle $W_\uout$ exceeds the work done on the oscillator $W_\uin$. Consitently, the heat obtained from the hot reservoir by the system $Q_\uh$ exceeds the heat released from the system to the cold reservoir $Q_\uc$.}
\label{fig:cycle}
\end{figure}

Perhaps the simplest model system of a QHE is a quantum harmonic oscillator (QHO) of variable frequency coupled to two heat reservoirs with peaked line shapes, so that the coupling strength to each of the reservoir can be varied by driving the oscillator in and out of resonance with the respective reservoir. Choosing the hot reservoir central frequency higher than the cold reservoir central frequency, a cycle depicted in Fig.~\ref{fig:cycle} can be achieved~\cite{qthermo, gemmer}. This is called the quantum Otto cycle~\cite{Kosloff}. Typically, in this scenario the frequency of the QHO is considered to be controlled by an external drive. There are plenty of examples of such quantum heat engines as cited above~\cite{rossnagel, vonL, vonH, Klatzow, deAssis, Peterson, Bouton}. In this case, however, the device is not autonomous, but completely dependent on constantly being controlled by some external force that in practical scenarios turns out to consume macroscopic amounts of power, greatly exceeding any work extracted by the heat engine from the thermal gradient. In this paper however, we propose a device, where the driving of the tunable QHO is incorporated into the device itself on the mesoscopic level, removing the need for macroscopic external driving. Our proposal, in essence, realizes an autonomous QHE, which continuously produces work once set to motion, without the need for any external control fields after the initialization phase.

In the spirit novel quantum technologies and inspired by the simple QHO model, we propose a way of utilizing systems realizing an optomechanical Hamiltonian~\cite{optomech, qoptom} as an autonomous quantum heat engine. In a prototypical optomechanical system, an optical cavity mode with frequency $\omega_{\ua}$ is coupled to a slower mechanical mode with frequency $\omega_{\textup{b}}$ so that the frequency of the optical mode is modulated by the mechanical displacement $x$, approximately as $\omega_{\ua}(x)\approx\omega_{\ua}^0+x\frac{\partial \omega_{\ua}}{\partial x}\big|_{x=0}$. Let the optical cavity be further coupled to two heat reservoirs with peaked spectral densities such that the center angular frequencies of the hot reservoir $\omega_{\uh}$ and cold reservoir the $\omega_{\uc}$ fulfill $\omega_{\uh} > \omega_{\ua}(x) > \omega_\uc$. Assigning temperatures $T_{\uh}$ and $T_{\uc}$ to the hot and cold reservoirs, respectively, such that $T_{\uh} > T_{\uc}$, leads to a device where the state of the optical cavity may undergo the quantum Otto cycle. The cavity mode at $\omega_{\ua}$ will traverse between the characteristic frequencies of the heat reservoirs driven by the mechanical mode. The optical mode will therefore pump photons from the high-frequency high-temperature environment to the low-frequency low-temperature environment, leading to energy boost to the mechanical mode owing to energy conservation. In this paper we aim to show, that the thermal energy released over the cycle is transformed, at least partially, into the amplitude  of the coherent state of the mechanical mode, thus available for useful work in the realm of quantum devices. For the sake of simplicity, we refer to the modes used in our proposal as the optical and the mechanical mode, but our theoretical proposal is hardware agnostic. If there are two modes and reservoirs that fulfill the discovered conditions, such a physical system is prone to be used as an autonomous quantum heat engine.

The proposed  device is a heat engine in the most fundamental sense---a heat engine is, by definition, a device that transforms heat into deterministic mechanical motion~\cite{Senft, Van_Wylen}. It should be noted, however, that this is not the first time an optomechanical system is used as the basis for a QHE. References~\cite{Dong, Zhang1, Zhang2} extensively analyze the possibility of realizing a coherently driven quantum Otto cycle in an optomechanical system. The above-mentioned references~\cite{Naseem2019, Izadyari2022, Hardal2017} utilize optomechanical systems, but rely on periodic incoherent thermal drives. Consequently, the coupling to the heat bath needs to be controlled by some external method, possibly consuming significant power. In contrast to such models, we propose a method which is fully free of temporally varied external controls and aim for a device that can work fully autonomously drawing power from constant-temperature heat reservoirs with static coupling spectra. Further, the references~\cite{Mari2015, Gelbwaser-Klimovsky2015, Gelbwaser-Klimovsky_2013} study theoretical models of autonomous QHEs based on optomechanical systems on the level of Markovian master equations. In general, it may be difficult to justify the Markov approximation if the heat reservoirs are spectrally separated and strongly structured in the relevant frequency range of the system, as illustrated in Fig.~\ref{spectra}. Thus, in contrast to the previous works, we do not invoke the Markov approximation, but rather consider the non-Markovian dynamics. Consequently, we rigorously take into account the spectral shape arising in all bosonic thermal sources due to quantum effects. As a result, our QHE exhibits autonomous cyclic dynamics in contrast to the traditionally classified completely continuous QHEs and externally controlled reciprocating QHEs.

This paper is organized as follows. In section~\ref{sec:simple}, we analyze the the proposed quantum heat engine by building an approximate analytical model for it. Subsequently, we find reasonable estimates for the net output power and efficiency of the device along with some physical intuition. Next, in section~\ref{sec:hle}, we develop a more detailed quasiclassical model based on Heisenberg--Langevin equations to describe the intricate dynamics of the QHE. We further outline how to numerically solve the resulting coupled non-linear stochastic integro-differential equations. In section~\ref{sec:dynamics}, we investigate the temporal evolution of the device according to the quasiclassical model. We obtain estimates for the power and efficiency, and analyze the power fluctuations inherently present in any QHE. Finally, in section~\ref{sec:conclusions}, we summarize our findings, draw conclusions, and provide our insight for the fruitful future research on this autonomous quantum heat engine.

\section{Analytical Model of the Quantum Heat Engine}
\label{sec:simple}

In order to gain physical intuition of the working mechanism of our device, we estimate the efficiency of the proposed QHE by an approximate analytical model. We begin with the usual model of the adiabatic quantum Otto cycle~\cite{Kosloff} and proceed by approximately 

\begin{figure}[!ht]
\centering
\includegraphics[width=0.6\textwidth]{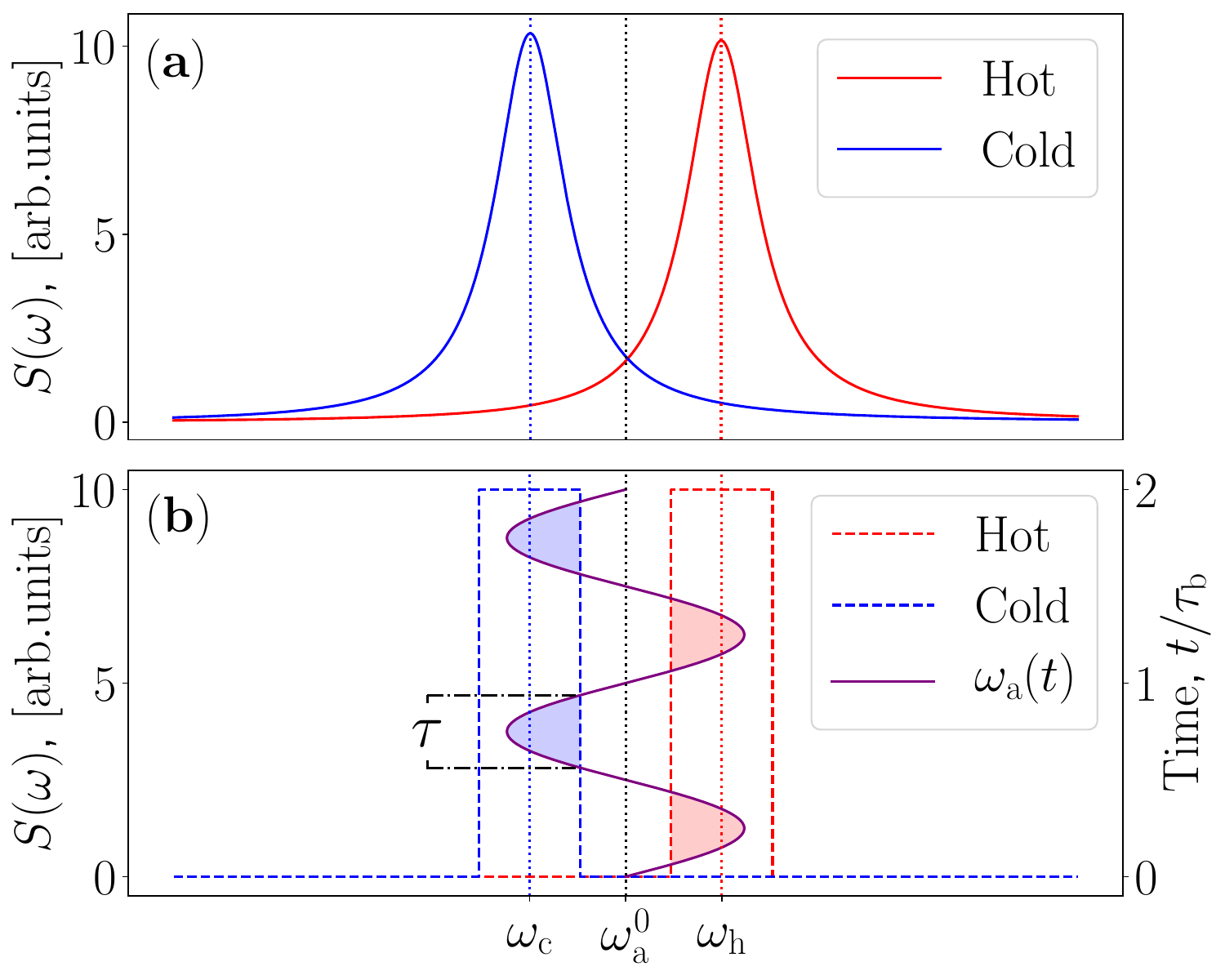}
\caption{\textbf{Spectral densities of the reservoirs.} (\textbf{a}) Experimentally realizable Lorentzian line shapes of the cold (blue color) and hot (red color) reservoirs. The vertical dotted lines denote the center angular frequencies of the cold and hot reservoirs, $\omega_\uc$ and $\omega_\uh$ respectively, and the unbiased angular frequency of the optical mode $\omega_\ua^0$. (\textbf{b}) Effective step function spectra denoted by the dashed lines. The modulation of the optical-mode frequency $\omega_\ua(t)$ in time is shown by the sinusoidal purple line. The shaded areas represent the interaction regions of the optical mode with the heat reservoirs. The interaction time $\tau$ is indicated in the figure and assumed equal for the two reservoirs.}
\label{spectra}
\end{figure}

\noindent taking into account the finite widths of the reservoir spectra. First, let us assume that the heat reservoirs have a step-function-like spectra rather than some experimentally encountered smooth peaked line shape, as depicted in Fig.~\ref{spectra}. In this case the coupling to each reservoir is sharply turned on and off without any residual coupling when the frequency of the optical mode moves into and out of the corresponding spectrum, respectively. The step function spectra are centered around $\omega_{\uh}$ and $\omega_\uc$, respectively, and are assumed to have equal widths $l_\uh=l_\uc$. Our optical mode frequency $\omega_\ua(t)$ oscillates about $\omega_\ua^0$, which is assumed to be centered in between $\omega_{\uh}$ and $\omega_\uc$. We assume that the optical mode frequency $\omega_\ua(t)$ traverses from one reservoir to the other periodically in time with the period $\tau_\ub=2\pi/\omega_\textup{b}$. Further, we define the interaction time $\tau$, which we assume to be equal for both reservoirs since the frequency $\omega_\ua(t)$ spends an equal amount of time per cycle in each frequency range defined by the step function of each reservoir. (See Fig.~\ref{spectra} for an illustration and Supplementary materials for theoretical details.) We further define the average angular frequency over the interaction period $\bar{\omega}_\ua^\textup{h/c}$, which is merely the time average of $\omega_\ua(t)$ over $\tau$, respectively for each reservoir. We use this average value as the effective angular frequency of the optical mode in the corresponding region. Note that we allow the possibility of $\omega_\ua(t)$ reaching values beyond the extents of the step function spectra.

To estimate the net work output of this QHE, we need to compute the mean number of photons transferred from the hot to the cold reservoir per cycle. Let the optical mode be coupled to each of the reservoirs so that the dissipation rates are $\Gamma_{\textup{h}}$ and $\Gamma_{\textup{c}}$, respectively. The mean number of thermal photons in the optical mode under the steady-state operation after the interaction period $\tau$ with the hot and cold reservoir is given, respectively by
\begin{align}
n_\uh&=N_{\uh}^{\textup{th}}-(N_{\uh}^{\textup{th}}-n_{\uc})\ue^{-\Gamma_{\uh}\tau},\\
n_\uc&=N_{\uc}^{\textup{tc}}-(N_{\uc}^{\textup{th}}-n_{\uh})\ue^{-\Gamma_{\uc}\tau},
\end{align}
where $N_{\textup{h/c}}^{\textup{th}}=1/\left(\exp{\frac{\hbar\bar{\omega}_\ua^\textup{h/c}}{k_\uB T_{\textup{h/c}}}}-1\right)$ is the mean photon number of the optical mode if it is fully thermalized with the hot or the cold reservoir, respectively, $\hbar$ is the reduced Planck constant, and $k_\uB$ is the Boltzmann constant. This set of equations can be solved for the steady state photon numbers given by
\begin{align}
n_{\uh}=\frac{N_{\uh}^{\textup{th}}\left(1-\ue^{-\Gamma_{\uh}\tau}\right)+N_{\uc}^{\textup{th}}\left(1-\ue^{-\Gamma_{\uc}\tau}\right)\ue^{-\Gamma_{\uh}\tau}}{1-\ue^{-(\Gamma_{\uh}+\Gamma_{\uc})\tau}},\\
n_{\uc}=\frac{N_{\uc}^{\textup{th}}\left(1-\ue^{-\Gamma_{\uc}\tau}\right)+N_{\uh}^{\textup{th}}\left(1-\ue^{-\Gamma_{\uh}\tau}\right)\ue^{-\Gamma_{\uc}\tau}}{1-\ue^{-(\Gamma_{\uh}+\Gamma_{\uc})\tau}}.
\label{n_th}
\end{align}
Note that we have used an equal interaction time $\tau$ for the reservoirs, without loss of generality since the only the product $\Gamma_{\textup{h/c}}\tau$ affects the end result, and hence different interaction times may be compensated by the dissipation rates. 

In this paper, we follow the traditional division of work and heat in quantum systems. Work is considered to be the energy related to changing the frequency of the optical mode operating here as the \emph{working fluid} and heat is considered to be the energy related to changing the population of the fluid. We assume that all the net energy released in a cycle from the reservoirs is transferred to the mechanical mode and that the photon number of the optical mode only changes during the interactions with the reservoirs. In this case, we may compute the average energy output per cycle and the average output power as 
\begin{align}
\label{ecyc}
E_{\textup{cyc}}&=\hbar(\Delta\bar{\omega}_{\ua} n_{\uh}-\Delta\bar{\omega}_{\ua} n_\uc)=\hbar\Delta\bar{\omega}_{\ua}\Delta n,\\
P&=\frac{\omega_\textup{b}}{2\pi}E_{\textup{cyc}},
\label{pow}
\end{align}
where $\Delta\bar{\omega}_{\ua}=\bar{\omega}_\ua^{\uh}-\bar{\omega}_\ua^\uc$ and $\Delta n$ is the difference of the mean photon number in the optical mode during a cycle. In the limit of ideal adiabatic operation, the efficiency of this device is defined in the usual way, which we find to coincide with the Otto efficiency~\cite{Kosloff} given by the effective frequencies as $\eta_\mathrm{eff}=\Delta W/\Delta Q_\uh=E_\mathrm{cyc}/(\hbar\bar{\omega}_\ua^\uh\Delta n)=1-\bar{\omega}_\ua^\uc/\bar{\omega}_\ua^\uh$. This efficiency is smaller than the often-used efficiency given by the peak-to-peak compression ratio $\kappa=(\omega_\ua^0+\Delta\omega_a/2)/(\omega_\ua^0-\Delta\omega_a/2)$: $\eta_\mathrm{eff}\leq\eta_\mathrm{max}=1-\kappa^{-1}$. 

\begin{figure}[!ht]
\centering
\includegraphics[width=1.0\textwidth]{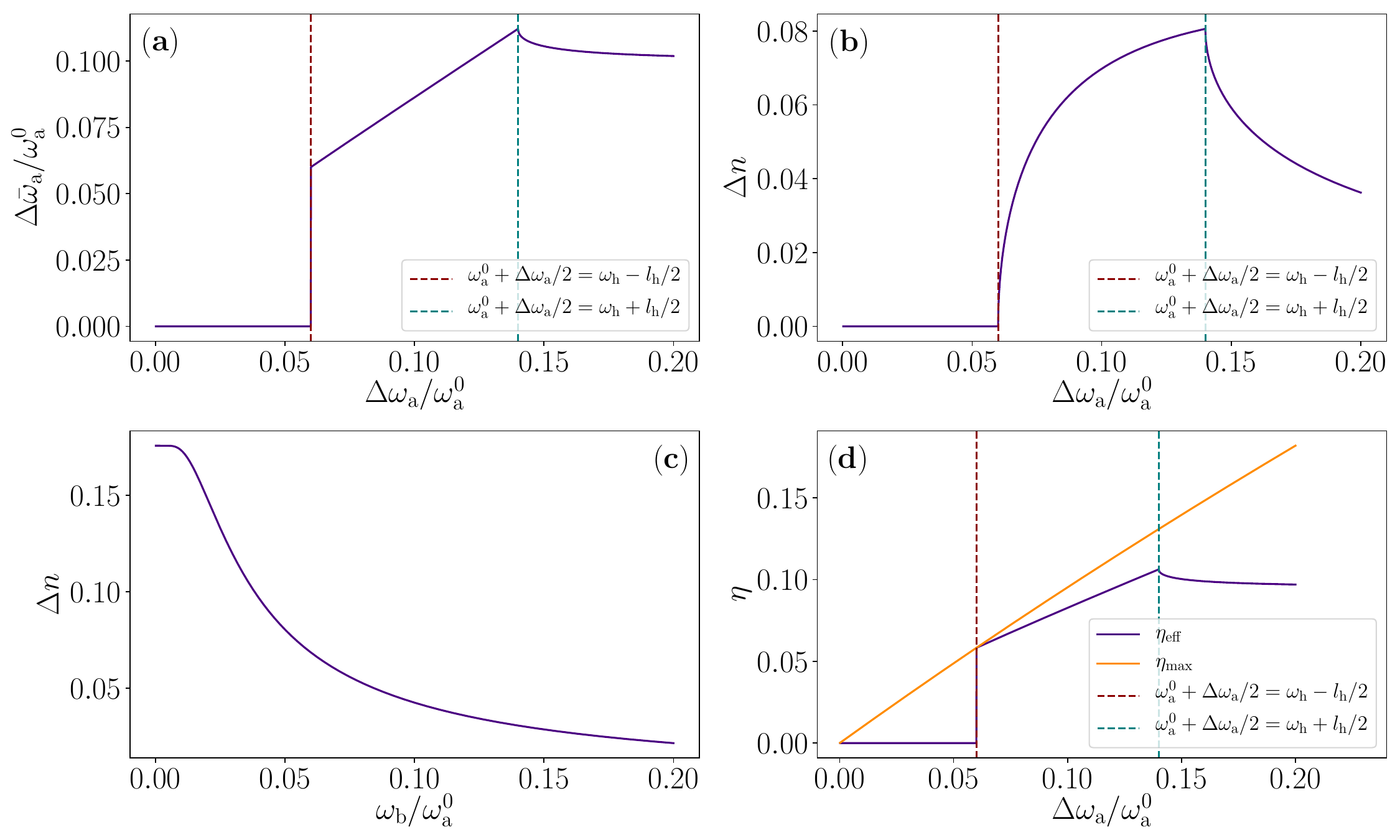}
\caption{\textbf{Derivative parameters and efficiency of quantum heat engine according to the analytical model.} (\textbf{a}) Difference of average frequencies over the interaction periods of the hot and cold reservoirs, $\Delta\bar{\omega}_\ua=\bar{\omega}_\ua^{\uh}-\bar{\omega}_\ua^\uc$, as a function of the peak-to-peak modulation amplitude $\Delta\omega_\ua$ of the optical-mode frequency $\omega_\ua(t)=\omega_\ua^0+\Delta\omega_\ua/2\sin(\omega_\ub t)$. (\textbf{b}, \textbf{c}) Difference of the mean photon number of the optical mode over the cycle as a function of (\textbf{b}) $\Delta\omega_\ua$ and (\textbf{c}) the mechanical mode frequency $\omega_\ub$. (\textbf{d}) Efficiency of the device $\eta_\mathrm{eff}=1-\bar{\omega}_\ua^\uc/\bar{\omega}_\ua^\uh$ as a function of $\Delta\omega_\ua$. 
The dashed vertical lines in (\textbf{a}), (\textbf{b}), and (\textbf{d}) indicate the threshold where the amplitude is just high enough for the optical-mode frequency to enter the non-vanishing regions of the step-function-like power spectra of the reservoirs (left, dark red) and the threshold where the frequency leaves at its extrema the non-vanishing spectra. The parameters used here are the following: $\Delta\omega_\ua = 0.139\times\omega_\ua^0$, $\omega_\ub=0.05\times\omega_\ua^0$, $k_\mathrm{B}T_\uh=0.56\times\hbar\omega_\ua^0$, $k_\mathrm{B}T_\uc=0.11\times\hbar\omega_\ua^0$, $\omega_\uh=1.03\times\omega_\ua^0$, $\omega_\uc= 0.97\times\omega_\ua^0$, $\Gamma_\uh=\Gamma_\uc=0.022\times\omega_\ua^0$, and  $l_\mathrm{h}=l_\mathrm{c}=0.04\times\omega_\ua$}
\label{fig:heuristic_params}
\end{figure}

\begin{figure}[!ht]
\centering
\includegraphics[width=\textwidth]{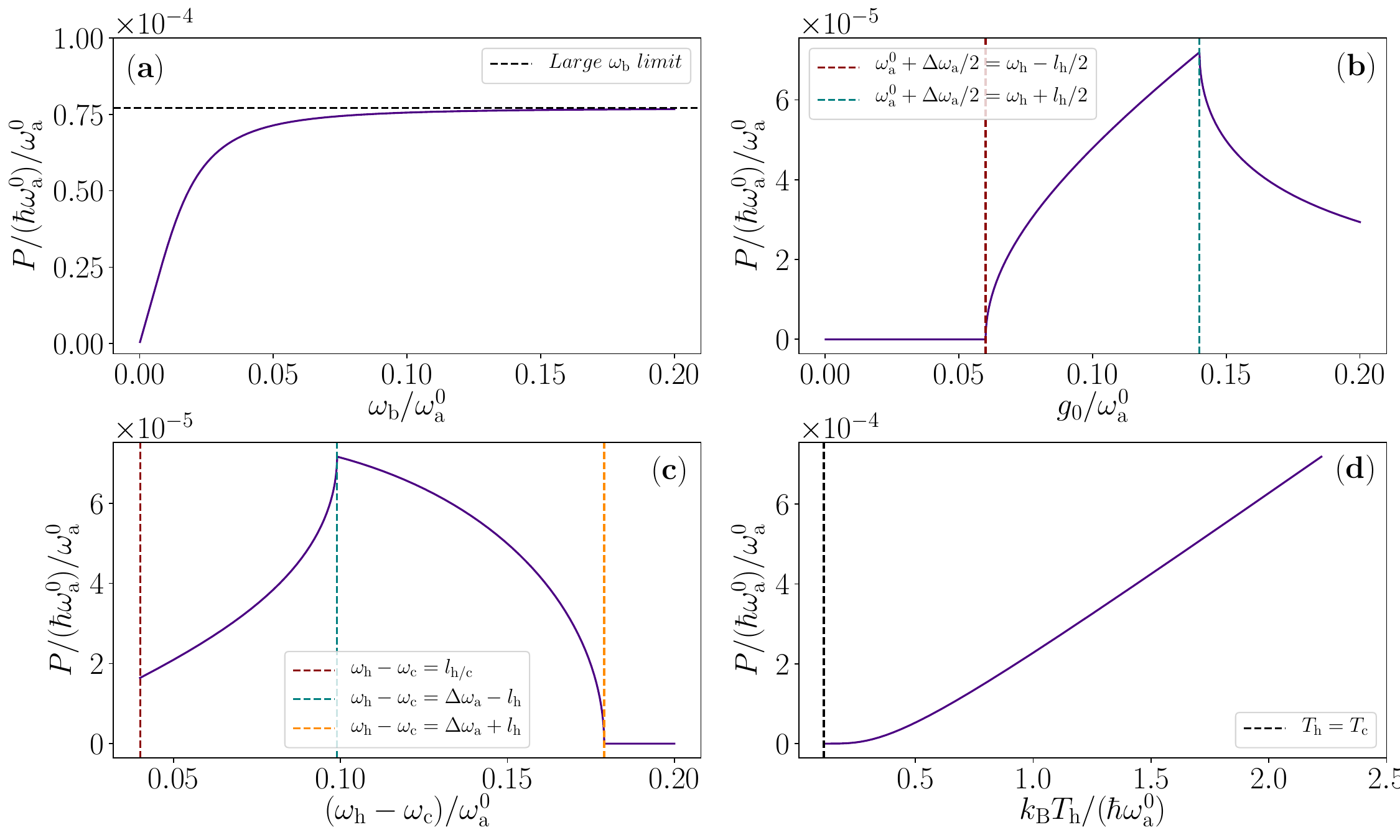}
\caption{\textbf{Output power of the quantum heat engine according to the analytical model.} (\textbf{a}--\textbf{d}) Output power as a function of (\textbf{a}) the mechanical mode angular frequency $\omega_\ub$, (\textbf{b}) the peak-to-peak modulation amplitude of the optical-mode angular frequency  $\Delta\omega_\ua$, (\textbf{c}) the difference between centre angular frequencies of the power spectra of the heat reservoirs $\omega_\uh-\omega_\uc$, and (\textbf{d}) the temperature of the hot reservoir $T_\uh$. The dashed vertical lines indicate the cases of zero distance (left, dark red), ideal distance (centre, turquoise), and too large distance (right, yellow) of the reservoir spectra with respect to the modulation amplitude $\Delta\omega_\ua$. We use identical parameters values to those in Fig.~\ref{fig:heuristic_params}.}
\label{fig:heuristic_p}
\end{figure}

This model already has a considerable number of adjustable parameters. We aim to illuminate the meaning and implications of the most important parameters here. In Fig.~\ref{fig:heuristic_params}, we show some derivative parameters of interest as a function of more fundamental, system parameters. In Fig.~\ref{fig:heuristic_p} we study the power in Eq.~\eqref{pow} and various limits as a function of the most interesting parameters. These graphs illustrate the dynamics and parameter ranges of the device quite well. Note that in realized optomechanical systems, the frequency of the mechanical mode is typically much lower than what is used here, but such a case will merely lead to closely adiabatic dynamics with respect to the thermalization and is theoretically easier and less interesting to analyze than the case studied here.

First, we observe from Fig.~\ref{fig:heuristic_p}(\textbf{a}) that the output power increases up to a limit with increasing $\omega_\ub$. This limit is a result of the fact that there is a trade-off between increasing speed of extracting $E_{\textup{cyc}}$ and decreasing the interaction time $\tau$ resulting in lower $\Delta n$ and hence lower $E_{\textup{cyc}}$ as shown in Fig.~\ref{fig:heuristic_params}(\textbf{c}). One of the most important fundamental parameters here is how much we vary $\omega_\ua(t)$ over the cycle. We define $\Delta\omega_\ua$ as the peak-to-peak modulation amplitude of $\omega_\ua(t)$, i.e., $\omega_\ua(t)$ will oscillate between $\omega_\ua^0-\Delta\omega_\ua/2$ and $\omega_\ua^0+\Delta\omega_\ua/2$. In Fig.~\ref{fig:heuristic_p}(\textbf{b}), we show the output power as a function of $\Delta\omega_\ua$. We clearly observe the limits of too low and too high modulation amplitudes, i.e., not reaching the reservoir at all or overshooting. This phenomenon is also observed in Figs.~\ref{fig:heuristic_params}(\textbf{a}) and~\ref{fig:heuristic_params}(\textbf{b}). In Fig.~\ref{fig:heuristic_p}(\textbf{c}) we see compatible behaviour as a function of the difference between the reservoir frequencies. Finally, in panel (\textbf{d}) of Fig.~\ref{fig:heuristic_params} we plot the efficiency of the heat engine. We note that utilizing the whole width of the step function reservoir in frequency gives the best efficiency, as expected. Increasing the peak-to-peak modulation amplitude and the energy gap between the reservoirs would increase the compression ratio, therefore allowing the increasing of the efficiency closer to the Carnot efficiency $\eta_\mathrm{C}=1-T_\uc/T_\uh\approx 80\%$.

To finalize our examination of the analytical model, we estimate the dissipation rate and the resulting quality factor of the mechanical mode required for steady-state operation of the QHE. Note that in practice, this effective dissipation of the mechanical mode is a result of using the power flowing into the mechanical mode in a desired use case. 

We assumed above that all the net thermal energy extracted over a cycle is transferred to the coherent motion of the mechanical mode. Under this assumption, which we justify below with a numerical model, we can match the power calculated above and the dissipated power from the mechanical mode in order to estimate the required quality factor of the mechanical mode. The average dissipated power of the mechanical mode can be expressed as $P_\ub=\Gamma_\ub \hbar n_\ub\omega_\ub$, where $\Gamma_\ub$ is the dissipation rate and $n_\ub$ the mean phonon number of the mechanical mode. Assuming that the mechanical mode modulates the optical mode frequency by $\Delta\omega_\ua\approx 2\sqrt{n_\ub} g_0$, where $g_0$ is the optomechanical coupling constant~\cite{optomech} discussed below, one can solve for the unknown parameter $n_\ub$. By setting the dissipated power equal to the power obtained in Eq.~\eqref{pow} and using Eq.~\eqref{ecyc}, we obtain $\Gamma_\ub=2\Delta\bar{\omega}_\ua\Delta n g_0^2/(\pi\Delta\omega_\ua^2)$.  We read the optimal values for $\Delta\omega_\ua$, $\Delta\bar{\omega}_\ua$, and $\Delta n$ from Fig.~\ref{fig:heuristic_p} and set the optomechanical coupling to a reasonable value, $g_0=0.01\times\omega_\ua^0$, to retrieve a numerical value of $\Gamma_\ub\approx 2.6\times 10^{-5}\times\omega_\ua^0$, which translates to a quality factor of about $Q_\ub=\omega_\ub/(2\Gamma_\ub)=960$ for the mechanical mode. Note this estimation does not depend on how the excitations are removed from the mechanical mode. We merely state that one can extract energy at this rate from the the mechanical mode. The extracted energy is interpreted as the work output of our heat engine. Note that, even though in reality there is also incoherent thermal occupation in the mechanical mode, we have not considered it here, nor does the analytical model capture the possibility. This will be discussed more in section~\ref{sec:dynamics}.

The analytical model presented here provides us intuition about the physics of the device and reasonable estimates on the output power and other parameters. It, however, assumes step-like power spectral densities of the baths and that the net thermal energy extracted from the reservoirs is fully transferred into the coherent motion of the mechanical mode. The analytical model does not allow us to investigate the dynamics of the whole system from arbitrary initial conditions. Of course, we also made a number of other rather crude approximations in order to simplify the calculations to the bare minimum. Below, we develop a more detailed theoretical description of the device allowing the investigation of the temporal evolution given arbitrary initial states, describing the optomechanical coupling more carefully and taking into account experimentally feasible spectral shapes of the reservoirs.

\section{Heisenberg--Langevin Equations for the Quantum Heat Engine}
\label{sec:hle}
\subsection{Derivation of the Equations of Motion}

As discussed above, we consider a system realizing an optomechanical Hamiltonian $\hat{H}_\textup{opt}$~\cite{optomech, qoptom} of an optical mode with bare angular frequency $\omega_\ua^0$ and a mechanical mode with angular frequency $\omega_\textup{b}$. The optical mode is coupled to two non-Markovian heat reservoirs~\cite{Chang} with peaked spectral line shapes at different central frequencies. The total Hamiltonian of the optomechanical system and its reservoirs is written as 
$$\hat{H}=\hat{H}_\textup{opt}+\hat{H}_{\uh}+\hat{H}_\uc+\hat{H}_\textup{int}^\uh+\hat{H}_\textup{int}^\uc,$$ 
where
\begin{subequations}
\begin{align}
\hat{H}_\textup{opt}&=\hbar\Big[\omega_\ua^0\hat{a}^{\dagger}\hat{a}+\omega_\textup{b}\hat{b}^{\dagger}\hat{b}-g_0\hat{a}^{\dagger}\hat{a}\left(\hat{b}^{\dagger}+\hat{b}\right)\Big],\\
\hat{H}_{\uh}&=\hbar\sum_k\omega_{\uh,k}\hat{h}_k^{\dagger}\hat{h}_k,\\
\hat{H}_{\uc}&=\hbar\sum_k\omega_{\uc,k}\hat{c}_k^{\dagger}\hat{c}_k,\\
\hat{H}_{\text{int}}^\uh&=\hbar\sum_k \lambda_{\uh,k}\left(\hat{a}^{\dagger}+\hat{a}\right)\left(\hat{h}_k^{\dagger}+\hat{h}_k\right),\\
\hat{H}_{\text{int}}^\uc&=\hbar\sum_k \lambda_{\uc,k}\left(\hat{a}^{\dagger}+\hat{a}\right)\left(\hat{c}_k^{\dagger}+\hat{c}_k\right),
\end{align}
\end{subequations}
where $\hat{a}$ and $\hat{b}$ are the annihilation operators of the optical and the reservoirs mode, respectively, $g_0$ is the  optomechanical coupling constant~\cite{optomech}, the operators $\hat{h}_k$ and $\hat{c}_k$ are the annihilation operators of the $k$:th modes at angular frequencies $\omega_{\uh,k}$ and $\omega_{\uc,k}$ in the hot and cold reservoirs, respectively, and the corresponding coupling constants between the reservoir modes and the optical mode are $\lambda_{\uh,k}$ and $\lambda_{\uc,k}$, respectively. Thus, $\hat{H}_{\uh}$ and $\hat{H}_{\uc}$ describe the reservoirs as infinite sums of bosonic modes. In addition, $\hat{H}_{\textup{int}}^{\uh}$ and $\hat{H}_{\textup{int}}^{\uc}$ describe the linear interaction between the reservoir modes and the optical mode.

The Heisenberg equations of motion for the optical and mechanical modes, as well as for the reservoir degrees of freedom are given by
\begin{subequations}
\begin{align}
\label{aeq}
\dot{\hat{a}}&=-\ui\omega_\ua^0\hat{a}+\ui g_{0}\hat{a}(\hat{b}^{\dagger}+\hat{b})-\ui\sum_k \lambda_{\uh, k}\left(\hat{h}_k^{\dagger}+\hat{h}_k\right)-\ui\sum_k \lambda_{\uc, k}\left(\hat{c}_k^{\dagger}+\hat{c}_{k}\right),\\
\label{beq}
\dot{\hat{b}}&=-\ui\omega_\textup{b}\hat{b}+\ui g_{0}\hat{a}^{\dagger}\hat{a},\\
\label{heq}
\dot{\hat{h}}_k&=-\ui\omega_{\uh, k}\hat{h}_{k}-\ui\lambda_{\uh, k}\left(\hat{a}^{\dagger}+\hat{a}\right),\\
\label{ceq}
\dot{\hat{c}}_k&=-\ui\omega_{\uc, k}\hat{c}_{k}-\ui\lambda_{\uc, k}\left(\hat{a}^{\dagger}+\hat{a}\right).
\end{align}
\end{subequations}
Let us integrate out the reservoir modes from the above set of equations. To this end, we solve Eqs.~\eqref{heq} and~\eqref{ceq} by explicit integration as
\begin{subequations}
\begin{align}
\hat{h}_k(t)&=\hat{h}_k(0)\ue^{-\ui\omega_{\uh, k}t}-\ui\lambda_{\uh, k}\int_0^{t} \ue^{-\ui\omega_{\uh, k}(t-\tau)}\left[\hat{a}^{\dagger}(\tau)+\hat{a}(\tau)\right]\ud\tau,\\
\hat{c}_k(t)&=\hat{c}_k(0)\ue^{-\ui\omega_{\uc, k}t}-\ui\lambda_{\uc, k}\int_0^{t} \ue^{-\ui\omega_{\uc, k}(t-\tau)}\left[\hat{a}^{\dagger}(\tau)+\hat{a}(\tau)\right]\ud\tau,
\end{align}
\end{subequations}
and substitute the solutions into equation \eqref{aeq}. Consequently, we arrive at the Heisenberg-Langevin equation (HLE),
\begin{align}
\nonumber
\dot{\hat{a}}&=-\ui\omega_\ua^0 \hat{a}+\ui g_{0}\hat{a}(\hat{b}^{\dagger}+\hat{b})-\hat{\xi}_\uh(t)-\hat{\xi}_\uc(t)\\
&\ \ \ \ +\int_0^{t} \left\{\mathcal{K}_\uh(t-\tau)+\mathcal{K}_\uc(t-\tau)\right\}\left[\hat{a}^{\dagger}(\tau)+\hat{a}(\tau)\right]\ud\tau,
\label{aeq_w_mk}
\end{align}
where we define the memory kernels
\begin{align}
\mathcal{K}_{\uh/\uc}(t-\tau)=2\ui\sum_k\lambda_{\uh/\uc,k}^2\sin[\omega_{\uh/\uc,k}(t-\tau)]=2\ui\int_0^{\infty} \mathcal{J}_{\uh/\uc}(\omega)\sin[\omega(t-\tau)]\ud\omega,
\label{memK}
\end{align}
with $\mathcal{J}_{\uh}(\omega)$ and $\mathcal{J}_{\uc}(\omega)$ being the spectral functions of the hot and cold reservoir, respectively, and the noise operators are given by
\begin{subequations}
\begin{align}
\hat{\xi}_{\uh}(t)&=\ui\sum_k \lambda_{\uh,k}\left[\hat{h}_k(0)\ue^{-i\omega_{\uh,k}t}+\hat{h}_k^{\dagger}(0)\ue^{i\omega_{\uh,k}t}\right],\\
\hat{\xi}_{\uc}(t)&=\ui\sum_k \lambda_{\uc,k}\left[\hat{c}_k(0)\ue^{-i\omega_{\uc,k}t}+\hat{c}_k^{\dagger}(0)\ue^{i\omega_{\uc,k}t}\right].
\end{align}
\label{noise}
\end{subequations}

At this point, the Markovian approximation~\cite{Chang, Dann, Rivas} is typically invoked such that the noise is replaced by white noise, characterized by $\expval{\hat{\xi}_{\uh/\uc}^{\dagger}(t)\hat{\xi}_{\uh/\uc}(t')}=\delta(t-t')$, where the average is taken over the reservoir degrees of freedom, and consequently the integrals in the above equations are reduced to linear dissipation~\cite{optomech, qoptom, Meystre}. However, as described above, the spectral density of the environment is here not frequency-independent, but rather, has a peaked line shape. Therefore, we consider the full non-Markovian model~\cite{Orieux2015}. The noise operators defined by Eqs.~\eqref{noise} have non-local temporal correlation relations, and together with the memory kernels, they characterize the non-Markovian features of the reservoirs and of the resulting dynamics of the system.

The coupled Eqs.~\eqref{beq} and~\eqref{aeq_w_mk} determine the dynamics of the optical and mechanical modes driven by the noise terms with damping arising from the memory kernel integrals. Unfortunately, these equations are challenging to solve exactly, even numerically, since they are non-linear operator equations of high dimension~\cite{optomech}. However, we can proceed by decomposing the operators into classical and quantum components, $\hat{a}=\alpha+\delta\hat{a}$ and $\hat{b}=\beta+\delta\hat{b}$, and linearize in terms of the quantum components~\cite{optomech, Zhang}. Namely, we assume the quantum components to provide a small correction to the classical dynamics. Consequently, we obtain 
\begin{subequations}
\begin{align}
\nonumber
\label{eq:alpha}
\dot{\alpha}&=-i\omega_\ua^0\alpha+ig_{0}\alpha(\beta^*+\beta)-\xi_\uh(t)-\xi_\uc(t)\\
&\ \ \ \ +\int_0^{t} \left\{\mathcal{K}_\uh(t-\tau)+\mathcal{K}_\uc(t-\tau)\right\}\left[\alpha^*(\tau)+\alpha(\tau)\right]\ud\tau,\\
\label{eq:beta}
\dot{\beta}&=-i\omega_\textup{b}\beta+ig_{0}\abs{\alpha}^2,\\
\label{eq:a_opp}
\delta\dot{\hat{a}}&=-i\omega_\ua'\delta\hat{a}+iG(\delta\hat{b}^{\dagger}+\delta\hat{b})+\int_0^{t} \left\{\mathcal{K}_\uh(t-\tau)+\mathcal{K}_\uc(t-\tau)\right\}\left[\delta\hat{a}^{\dagger}(\tau)+\delta\hat{a}(\tau)\right]\ud\tau,\\
\label{eq:b_opp}
\delta\dot{\hat{b}}&=-i\omega_\textup{b}\delta\hat{b}+i\left(G\delta\hat{a}^{\dagger}+G^*\delta\hat{a}\right),
\end{align}
\end{subequations}
where $G=g_{0}\alpha(t)$, we have transformed the noise operators $\hat{\xi}_{\uh/\uc}$ into stochastic complex-valued variables ${\xi}_{\uh/\uc}$, and $\omega_\ua'=\omega_\ua^0-g_0\left[\beta(t)+\beta^*(t)\right]$. These equations are exact up to the linearization and describe the dynamics of the system under the influence of the reservoirs. The quasiclassical equations \eqref{eq:alpha} and \eqref{eq:beta} retain the inherent non-linearity of the system, whereas the quantum equations \eqref{eq:a_opp} and \eqref{eq:b_opp} yield the quantum corrections. The classical equations can be understood as a mean-field-like approximation to the full quantum equations \eqref{aeq_w_mk} and \eqref{beq}. Thus, they may provide a reasonable approximate solution in many physical situations. For instance, in reference~\cite{Hardal2017}, where a QHE based on optomechanical system pumped by periodic incoherent drive is studied, the difference between quantum and classical models is found to be nighly negligible.

\subsection{Considerations for Numerical Solution of the Equations}

To analyze the system dynamics based on the approximate Heisenberg--Langevin equations derived in previous section, we resort to numerics. In order to relax the requirements of the computational resources, let us investigate a quasiclassical~\cite{Schmid1982} model based on Eqs.~\eqref{eq:alpha} and~\eqref{eq:beta}. Further, we are mostly interested in the non-linear dynamics of the system, captured by the classical equation, and have already made the assumption that the classical component of the dynamics is dominating. Using noise arising from thermal quantum fluctuations, the quantum statistics of the reservoirs is taken into consideration and a quasiclassical model for the system dynamics is achieved. Although we neglect the quantum fluctuations of the system itself~\cite{Paladino}, we consider the model to likely be accurate enough especially for average quantities since quasiclassical models similar to this have been shown to produce valuable insight into quantum systems interacting with heat sources~\cite{Schmid1982}.

Here, we numerically solve the coupled equations~\ref{eq:alpha} and~\ref{eq:beta} in the time domain. Strictly speaking, stochastic differential equations (SDEs), and more importantly the numerical solution methods, are only defined for Wiener processes, i.e. white noise~\cite{kloeden, Burrage}. Physicist and engineers have however stretched the limits of mathematical definitions, and considered equations with coloured noise obtaining physically meaningful results~\cite{kloeden, Pardoux, Adomian}. Therefore, we follow these footsteps and proceed with our analysis assuming that in the sense of small enough time-steps our noise is locally white. We interpret our SDE in the sense of Stratonovich stochastic calculus and apply a second-order predictor-corrector method~\cite{Burrage, kloeden, Day} to solve the above stochastic integro-differential equation~\cite{singh}. The coloured noise is generated from the power spectral densities, given below by Eq.~\ref{Svv}, by filtering zero-mean Gaussian white noise with a filter, the frequency response function $\mathrm{FRF}_{\uh/\uc}(\omega)$ of which satisfies $\abs{\mathrm{FRF}_{\uh/\uc}(\omega)}^2=S_{\uh/\uc}(\omega)$. Note also that since we are dealing with a stochastic system, single realizations of the dynamics exhibit random behaviour with little conclusions to be drawn. Thus, we simulate the system with a large number of repetitions and compute ensemble averages of the quantities of interest.

Let us specify the reservoir spectra and the temperature dependence of the heat reservoirs. First, we consider the following power spectral density of the thermal fluctuations~\cite{Devoret, Kubo} in the reservoir:
\begin{align}
S_\mathrm{h/c}(\omega)=\mathcal{F}\left[\expval{\hat{\xi}_\mathrm{h/c}^{\dagger}(t)\hat{\xi}_\mathrm{h/c}(t')}\right](\omega)=\mathcal{J}_\mathrm{h/c}(\omega)\coth(\frac{\hbar\omega}{2k_\uB T_\mathrm{h/c}}),
\label{Svv}
\end{align}
where $\mathcal{F}\left[f(t)\right](\omega)$ denotes the Fourier transform of function $f(t)$ and 
\begin{align}
\mathcal{J}_\mathrm{h/c}(\omega)=g_\mathrm{h/c}\frac{\hbar\omega^3}{\omega^2+\gamma_\mathrm{h/c}^{-2}(\omega^2-\omega_\mathrm{h/c}^2)^2},
\label{spec_dens}
\end{align}
where $g_\mathrm{h/c}$ is a dimensionless constant scaling the coupling strength between the optical mode and the corresponding reservoir, $\gamma_\mathrm{h/c}$ equals the full width of the spectrum at half maximum, $\omega_\mathrm{h/c}$ is the renormalized centre frequency, and $T_\mathrm{h/c}$ is the temperature of the reservoir as defined above. The hyperbolic cotangent term in the above expression arises by directly calculating the variance of noise operators following the usual analysis of generalized bosonic heat reservoirs~\cite{Weiss, zoller, Devoret}. The line shape given by $\mathcal{J}_\mathrm{h/c}(\omega)$ is chosen merely to consider spectra which are peaked in a reasonable way without an attempt to optimize them for output power. 

Let us make a minor remark about the parameters. We introduce two new parameter here, namely $g_{\uh/\uc}$ and $\gamma_{\uh/\uc}$. Unfortunately, these are not directly quantitatively comparable to their counterparts in the analytical model. Where $\Gamma_{\uh/\uc}$ in the analytical model is the relaxation time constant of the optical mode coupled to the reservoir, $g_{\uh/\uc}$ describes the coupling strength with the reservoir. These are closely related, but not the identical, and can vary from physical realization to another. The full width at half maximum of the spectrum in Eq.~\eqref{spec_dens} is given by $\gamma_{\uh/\uc}$. This can be compared to the width of the step function spectrum, $l_{\uh/\uc}$, more or less directly, keeping in mind that the spectrum here is smooth with polynomially decreasing tails, in stark contrast with the step function. (See Fig.~\ref{spectra} for comparison.)

\section{Dynamics and Performance of the Quantum Heat Engine}
\label{sec:dynamics}
\subsection{Temporal Evolution and Output Power}
\label{sec:temporal_evol}

\begin{figure}[!ht]
\centering
\includegraphics[width=\textwidth, trim={0, 0, 0, 0}, clip]{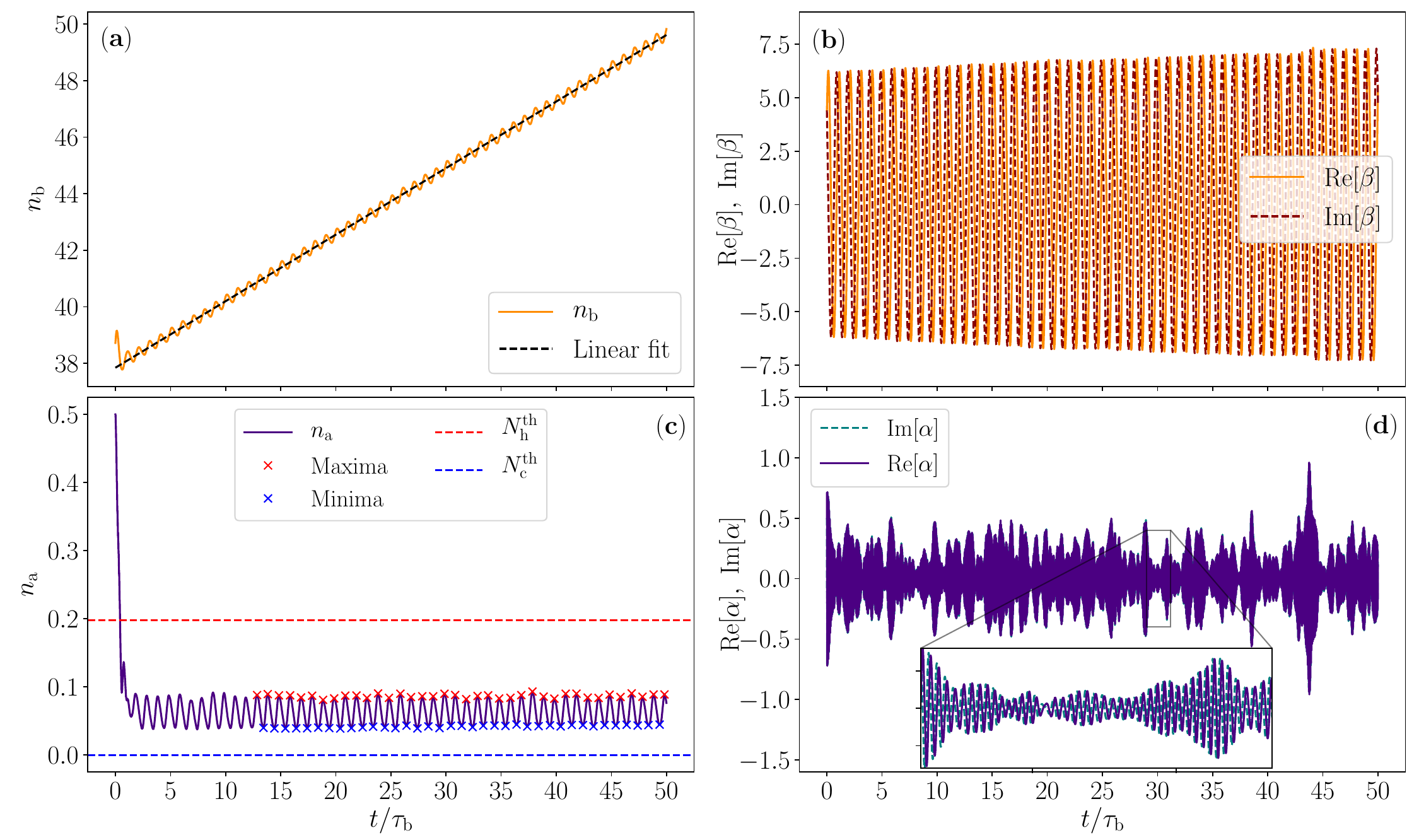}
\caption{\textbf{Temporal evolution of the quantum heat engine according to the quasiclassical Heisenberg--Langevin equations~\eqref{eq:alpha} and~\eqref{eq:beta}.} (\textbf{a})~Mean occupation number of the mechanical mode as a function of time. The dashed line shows a linear fit to the data in order to find out the average rate of occupation generation. (\textbf{b}) Single realization of the real and imaginary parts of the complex-valued amplitude of the mechanical mode as functions of time. (\textbf{c}) Mean occupation number of the optical mode as a function of time. We show the thermal photon numbers for the optical mode, $N_\uh^{\mathrm{th}}$ (red dashed line) and $N_\uc^{\mathrm{th}}$ (blue dashed line), given by the Bose--Einstein distribution at the reservoir temperatures. The red and blue crosses indicate the maxima and minima of the optical-mode occupation in each cycle after the transient. (\textbf{d}) Single realization of the complex-valued amplitude of the optical mode. The parameters used are $\omega_\textup{b}=0.048\times\omega_\ua^0$, $\omega_{\uh}=1.04\times\omega_\ua^0$, $\omega_\uc=0.964\times\omega_\ua^0$, $k_\mathrm{B}T_\uh=0.56\times\hbar\omega_\ua^0$, $k_\mathrm{B}T_\uc=0.11\times\hbar\omega_\ua^0$, $g_0=0.012\times\omega_\ua^0$, $g_\uh\approx 0.007$, $g_\uc\approx 0.0082$, $\gamma_\uh\approx 0.031\times\omega_\ua^0$, $\gamma_\uc\approx 0.025\times\omega_\ua^0$. At time instant $t=0$, the modes are initialized into coherent states corresponding to $n_\ua=0.5$ and $n_\ub=39$. The mean values are found by averaging over 1000 realizations of the noise.}
\label{time_evol}
\end{figure}

In Fig.~\ref{time_evol}, we show the temporal evolution of the amplitudes $\alpha$ and $\beta$ and excitation numbers $n_\ua=\langle|\alpha|^2\rangle$ and $n_\ub=\langle|\beta|^2\rangle$ of the optical and the mechanical mode, respectively, assuming negligible dissipation on the mechanical mode. Here, the average is an ensemble average with respect to many independent realizations of the noise trajectories $\xi_\uh(t)$ and $\xi_\uc(t)$. This theoretical scenario allows us to gain insight into the operation principle of the quantum heat engine. The parameters are chosen to roughly optimize the power generation in the mechanical mode with a constant temperature difference between the reservoirs. Full systematic optimization of the parameters, being somewhat computationally intensive, is left for future work.

Initially, the modes are set to coherent states with mean excitation numbers of $n_\ua=0.5$ and $n_\ub=39$, correspondingly. Firstly and most importantly, we observe from Fig.~\ref{time_evol}(\textbf{b}) that the initially prepared coherent state tends to be preserved in the evolution and even increasing its amplitude rather than winding towards a thermal state. The increasing amplitude of the oscillation is in agreement with Fig.~\ref{time_evol}(\textbf{a}) where the average excitation number of the mechanical mode steadily increases in time. These key observations justify the assumption we introduced for the analytical model that the net thermal energy extracted from the heat baths tends to convert into the coherent motion of the mechanical mode.

From the results of Fig.~\ref{time_evol}(\textbf{c}), we observe for the optical mode that instead of thermalizing to a steady state with a constant average photon number between the thermal occupation numbers $N_\uh^{\mathrm{th}}$ and $N_\uc^{\mathrm{th}}$ given by the Bose--Einstein distribution at the reservoir temperatures, the average photon number of the optical mode exhibits clear temporal oscillations. These oscillations take place between the boundaries set by the thermal photon numbers with a period given by the frequency of the mechanical mode $\omega_\ub$. This behaviour arises from driving the frequency of the optical mode by the mechanical mode between the peaked spectra of the hot and cold reservoirs that inject and absorb optical photons, respectively. In stark contrast to the traditional externally driven quantum heat engines however, driving is here purely caused by the internal dynamics of the device. In a qualitative agreement with our analytical model and Fig.~\ref{fig:cycle}, the average excitation number of  the mechanical mode exhibits oscillations phase shifted by $\pi/2$ from those of the optical mode, i.e., when the optical mode has a peak in its average photon number, the mechanical mode is gaining energy at maximum speed from the optical mode. 

Since the dissipation of the mechanical mode is neglected and the mechanical mode has a reasonably low amplitude, we do not observe the mechanical mode to stabilize into a steady state. However, we can estimate the upper limit of power, at which the mechanical mode accumulates energy. In Fig.~\ref{time_evol}(\textbf{a}), we employ a linear fit on the average excitation number of the mechanical and find that based on the slope of the fitted line, this device yields an average power of $P=8.63\times 10^{-5}\hbar(\omega_\ua^0)^2$, which is clearly higher than the power achieved according to the analytical model in section~\ref{sec:simple}. This is partially attributed to the slightly different parameter values and approximations used in the two models, but the main reason for observing a higher power output is, however, expected to be the initial thermalization transient experienced by the mechanical mode. From Figs.~\ref{time_evol}(\textbf{c}) and~\ref{time_evol}(\textbf{d}), it is clear that the optical mode reaches a state dominated by thermal noise very quickly, but the saturation of the mechanical mode may take significantly more time due to its remarkably weaker coupling to the thermal sources. Of course, we do not even wish the mechanical mode to reach a thermal state, but to retain its coherence as well as possible. 

A reasonable estimate for the average thermal occupation in the mechanical mode after a long time is given by the weighted average $N_\ub^\mathrm{th}=(g_\uh N_\ub^\uh+g_\uc N_\ub^\uc)/(g_\uh+g_\uc)\approx 6.1$, where $N_\ub^{\uh/\uc}$ is the mechanical-mode occupation fully thermalized at the reservoir temperature $T_{\uh/\uc}$ given by the Bose--Einstein distribution. This further supports the observation that increasing occupation well above $n_\ub=40$ must be at least partially coherent generation, as we would expect from Figs.~\ref{time_evol}(\textbf{b}) and~\ref{time_evol}(\textbf{d}). Therefore, it is reasonable to assume that the increasing occupation is partially due to coherent generation and partially due to thermalization. Let us next study the steady states of the mechanical mode with finite dissipation and after a long evolution in order to shed some light on the matter. 

\begin{figure}[ht]
\centering
\includegraphics[width=\textwidth, trim={0 0 0 0}, clip]{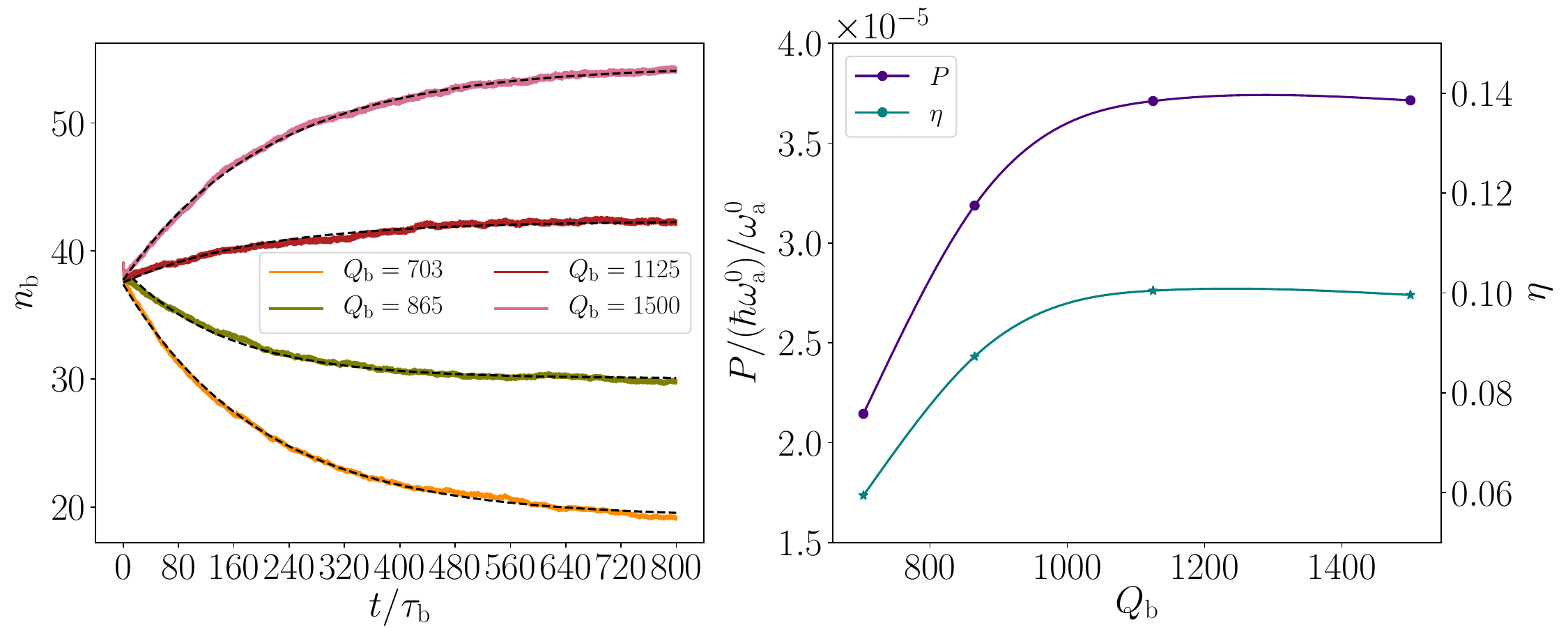}
\caption{\textbf{Operation of the quantum heat engine with finite extracted power.} (\textbf{a}) Temporal evolution of the average excitation number of the mechanical mode under different levels of dissipation, as indicated by the given quality factors. The dashed black lines show the fitted exponential evolution. (\textbf{b}) Output power (left vertical axis) and efficiency (right vertical axis) of the QHE as a function different levels of excess dissipation. The dots are the calculated values based on the simulation results extracted at the different levels of dissipation found in panel (\textbf{a}). The dots are connected by a solid line based on cubic spline interpolation. Details of the calculation can be found in the main text. The parameters and initial state used in the calculations are identical to those in Fig.~\ref{time_evol}.}
\label{decay}
\end{figure}

In Fig.~\ref{decay}(\textbf{a}), we show the temporal evolution of the average excitation number of the mechanical mode for different levels of excess dissipation, i.e., non-vanishing quality factors. As expected in this case, the excitation number exponentially reaches a steady state where we constantly extract work from the mechanical mode. Note, that the work here is simply dissipated away, but in practice it can be released to a waveguide and measured. From Fig.~\ref{decay}(\textbf{a}), we note that the quality factors of the mechanical mode are on a similar level to those found with the analytical model above.

Let us estimate the dissipated power based on the dissipation rate by the method used above in the analytical model. However, in order to separate between work and heat in the mechanical mode, we make the reasonable assumption that after a long time the mechanical mode stabilizes into a thermal coherent state~\cite{Oz1991}, and use the fact that the average occupation number of the thermal coherent state is given as a sum of the coherent and thermal occupations. We treat the average occupation number resulting from our simulation, $n_\ub$, as the total occupation number of the thermal coherent state, and use the above-obtained estimate for the thermal contribution $N_\ub^\mathrm{th}$. Thus the coherent contribution is given by $n_\ub^\mathrm{coh}=n_\ub-N_\ub^\mathrm{th}$, which allows us to interpret the coherent dissipated power as work and compute the corresponding power output as $P=\Gamma_\ub\hbar n_\ub^\mathrm{coh}\omega_\ub$. We show the obtained power in Fig.~\ref{decay}(\textbf{b}) for different levels of excess dissipation. We point out that for the highest efficiency with the mechanical-mode quality factor $Q_\ub=1125$, the thermal contribution in the mechanical mode occupation is only about $N_\ub^\mathrm{th}/n_\ub=14\%$ of the total occupation number. Furthermore, the power output obtained here is less than half of the upper-bound estimate obtained above. This observation supports the hypotheses of the thermalization transient. 

In addition, we find  in Fig.~\ref{decay}(\textbf{b}) that we can vary the dissipation level quite significantly without much disturbing the power generation. As long as the dissipation is on a reasonable level compared with the power generation rate found above, the systems seems to be able to find a steady state. Bear in mind that this is a dynamic process involving non-linear back and forth coupling between the mechanical and the optical mode as a function of the occupation number $n_\ub$. As opposed to the analytical model, this quasiclassical model takes into account the intricate dynamics owing to the optomechanical coupling.

\subsection{Efficiency}

From the above dynamics of our quantum heat engine, it is somewhat challenging to rigorously define and determine certain quantities traditionally studied in quantum heat engines, stemming from the lack of the external driving field. Typically, the work done on the working fluid by the external drive plays an important role in defining these quantities. As already mentioned in section~\ref{sec:simple}, work is the energy related to changing the Hamiltonian, or frequency, of the working fluid whereas heat is the energy related to changing the populations. The issue arising from the lacking external drive is somewhat more pronounced in the quasiclassical model than in the analytical model since for the quasiclassical model we are truly solving the dynamics of the whole optomechanical system, whereas in the analytical model, we still treat them as separate systems. 

Let us, nevertheless, try to estimate the efficiency of our quantum heat engine by using the method we utilized for the analytical model. We consider the optical mode to be the working fluid and assume that the thermal occupation of the  mechanical mode does not significantly vary over a cycle once the steady state is reached. We further approximate that the oscillation in the occupation of the optical mode is almost solely due to the interactions with the thermal reservoirs. This is a very good approximation, since the mechanical mode varies the occupation of the optical mode very little over a cycle due to the relatively low optomechanical coupling, $g_0\ll\omega_\ua^0$, and excitation level of the mechanical mode. (See Supplementary Fig.~\ref{fig:s6}\ S(1).) In this scenario, we find the heat absorbed by the device from the hot reservoir per cycle, $\Delta Q_\uh$, by finding the average variation of the photon number in the optical mode over cycle. This can be computed from the maxima and minima of the photon number, $n_\ua$, over many cycles, as depicted by the red and blue crosses in Fig.~\ref{time_evol}(\textbf{c}). The average work per cycle is simply given by the power: $\Delta W=P\tau_\ub$. With this method we estimate the efficiency, $\eta$, shown in Fig.~\ref{decay}(\textbf{b}) as a function of excess dissipation levels of the mechanical mode. We note that the efficiency is qualitatively remarkably close to the one achieved in the analytical model, which would suggest that the analytical model captures the essence of the physics relatively well. As one may expect, the efficiency is far below the Carnot efficiency, $\eta_\mathrm{C}\approx 80\%$, and also significantly below the Otto efficiency given by the compression ratio $\eta_\mathrm{max}\approx 16\%$.


Finally, we highlight an important detail, i.e., we cannot claim that the heating and cooling processes of the working fluid are perfectly isochoric, nor can we say that the compression and decompression are fully adiabatic. If we take into account the slowly vanishing tails of the reservoir spectra, it becomes evident that there is constant heat exchange between the optical mode and the reservoirs over the cycle. These considerations inevitably lead to the fact that we cannot, strictly speaking, call this cycle an Otto cycle in its ideal form. The cycle of our device displays more continuous dynamics, not perfectly split into well defined phases. Moreover, we point out that the above discussion considers only the internal efficiency of the device. In an experimental realization, there is also direct heat leakage from the hot reservoir to the cold reservoir through spurious parallel channels to the quantum heat engine, decreasing the total efficiency in practice.

\subsection{Stability and Power Fluctuations}

\begin{figure}[!ht]
\centering
\includegraphics[width=1\textwidth, trim={0 0 0 0}, clip]{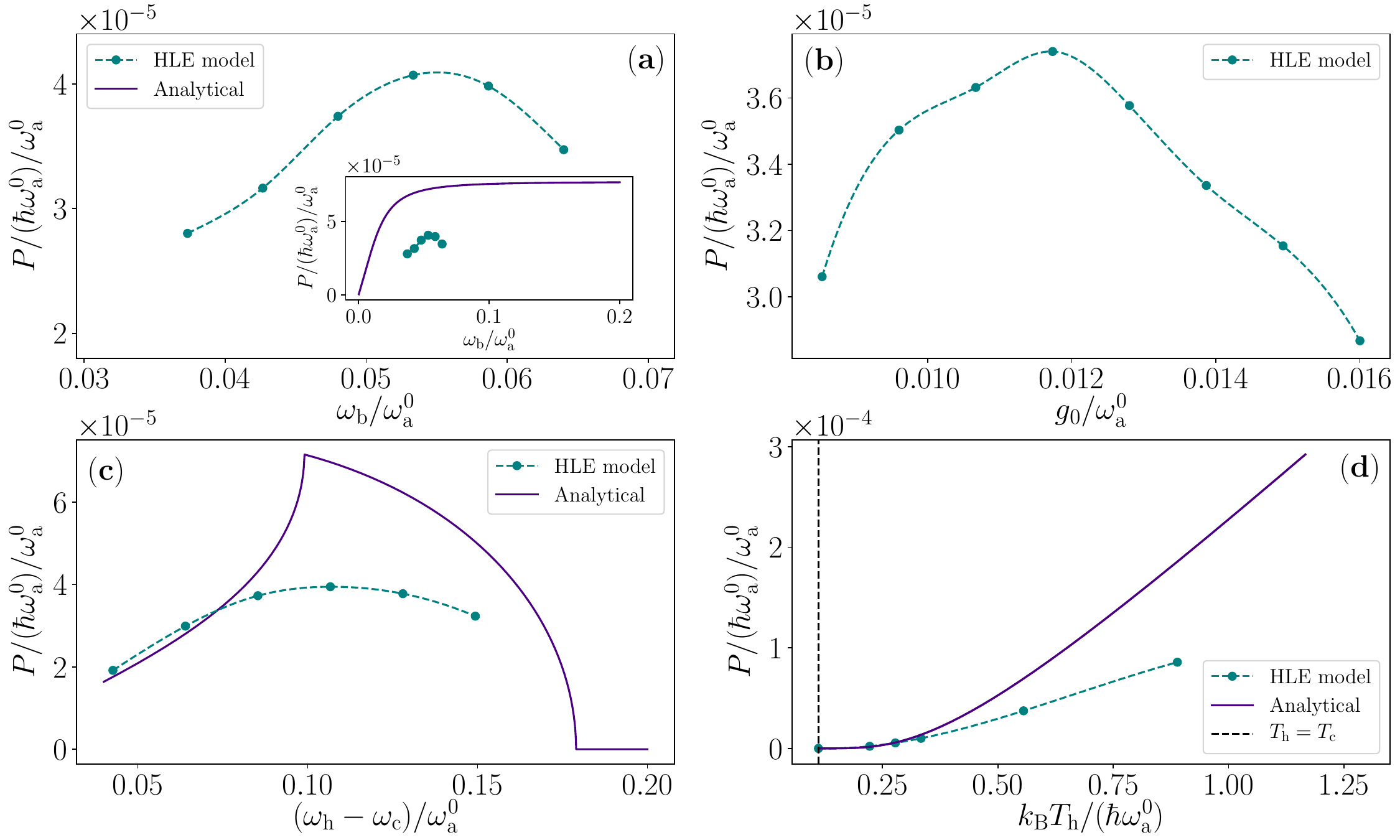}
\caption{\textbf{Output power of the quantum heat engine according to the quasiclassical model compared with the analytical model.} (\textbf{a}–\textbf{d}) Output power of the quasiclassical (cyan color) and the analytical (purple color) model as a function of (\textbf{a}) the mechanical mode angular frequency $\omega_\ub$, (\textbf{b}) the optomechanical coupling $g_0$, (\textbf{c}) the difference between the centre angular frequencies of the power spectra of the heat reservoirs $\omega_\uh-\omega_\uc$, and (\textbf{d}) the temperature of the hot reservoir $T_\uh$. For the quasiclassical model, we use identical parameters values to those in Fig.~\ref{decay} for $Q_\ub=1250$. The data for the analytical model is identical to those in Fig.~\ref{fig:heuristic_p}. The inset in (\textbf{a}) shows the output power in the full range of Fig.~\ref{fig:heuristic_p}(\textbf{a}). We use cubic spline method to interpolate (dashed cyan color) the numerical data (cyan color dots) of the Heisenberg--Langevin equation model. Note that because of the differences in the models and their parameters, we do not expect the models to accurately quantitatively agree.}
\label{fig:param_sweeps}
\end{figure}

In this section, we briefly analyze the ensemble fluctuations of the output power and the stability of the device with respect to some parameters. In Fig.~\ref{fig:param_sweeps}, we compare the net output power obtained from the quasiclassical model with the analytical model of section~\ref{sec:simple} as a function of a few of the most important parameters. Interestingly, there is a relatively narrow range of optimal mechanical-mode frequencies in the quasiclassical model depicted in Fig.~\ref{fig:param_sweeps}(\textbf{a}). In the analytical model, we do not restrict the mechanical-mode frequency in any way, and seems that it can be increased indefinitely, whereas in the quasiclassical model we clearly find a sweet spot for the frequency $\omega_\ub$. This discrepancy may be a result of the fixed peak-to-peak modulation amplitude $\Delta\omega_\ua$ assumed in the analytical model, whereas the amplitude in the quasiclassical model is a result of the temporal evolution determined by the other parameters.

In Fig.~\ref{fig:param_sweeps}(\textbf{c}), we find that the two models behave qualitatively similarly as a function of the difference between the centre angular frequencies of the reservoir power spectra. Note that the curve of the quasiclassical model does not display any sharp cusp as observed for the analytical model, owing to the continuous spectral functions in the quasiclassical model. The behaviour of the output power as a function of the hot-reservoir temperature is found in Fig.~\ref{fig:param_sweeps}(\textbf{d}) qualitatively quite similar to the analytical model, as expected. However, the analytical model predicts a steeper increase in the power as function of increasing hot-reservoir temperature than the HLE model, which is attributed to the accordingly rising mechanical-mode thermal occupation in the HLE model, which does not contribute to the work.

\begin{figure}[!ht]
\centering
\includegraphics[width=\textwidth, trim={0 0 0 0}, clip]{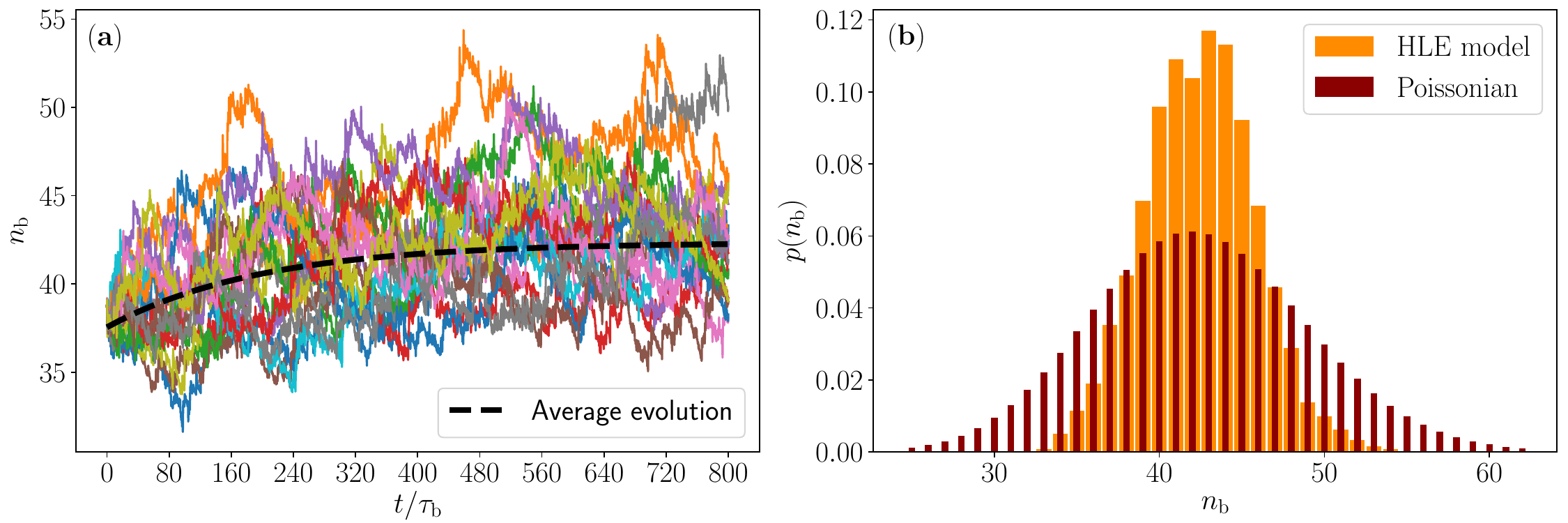}
\caption{\textbf{Mechanical mode occupation fluctuations of the quantum heat engine (QHE).} (\textbf{a}) Hundred realizations of the temporal evolution of the average excitation number of the mechanical mode under dissipation characterized by $Q_\ub=1125$. The black dashed line shows the average exponential behaviour as found in Fig.~\ref{decay}(\textbf{a}). (\textbf{b}) Probability distribution of the mechanical mode occupation number after reaching a steady state together with a Poisson distribution for reference. The parameters and initial state used in the calculations are identical to those in Fig.~\ref{time_evol}.}
\label{fig:fluctuations}
\end{figure}

It is interesting to investigate the relative occupation number fluctuations of the mechanical mode, because it gives insight to the excitation counting statistics one would observe when measuring the mechanical mode. Further, because the output power is directly related on the occupation number, as defined above through dissipated power, the relative output power fluctuations of the QHE~\cite{Batalhao, Barker} are consequently directly related to the occupation number fluctuations. In Fig.~\ref{fig:fluctuations}(\textbf{a}) we illustrate the occupation number fluctuations by showing hundred different realization of the temporal evolution of the excitation number of the mechanical mode under excess dissipation characterized by the quality factor $Q_\ub=1125$. In Fig.~\ref{fig:fluctuations}(\textbf{b}), we display the probability density for the occupation number after reaching a steady state and compare it to the Poisson distribution. We use the Fano factor~\cite{Fano}, which quantifies the deviation from a Poisson distribution, as a measure to compare the occupation number distribution to the Poissonian. The Fano factor is defined as $F=\sigma_P^2/\expval{P}$, where $\sigma_P^2$ and $\expval{P}$ are the variance and the expectation value of the occupation number, respectively. For the data in Fig.~\ref{fig:fluctuations}(\textbf{b}), we find $F=0.27$, which signifies far sub-Poissonian ($F<1$) statistics of the occupation number fluctuations. This suggest a smaller output power fluctuations for the QHE than for a Poisson process.

\section{Conclusions and Discussion}
\label{sec:conclusions}

We have proposed and analyzed an autonomous quantum heat engine in the spirit of an Otto cycle, where one harmonic mode, the working fluid, traverses the cycle driven by another harmonic mode. We utilized an optomechanical Hamiltonian for this realization and consequently referred to the working fluid as the optical mode and the driving mode as the mechanical mode. However, any physical system realizing the optomechanical Hamiltonian in the appropriate parameter range will suffice our heat engine. 

The feasible parameter range for the quantum heat engine is straightforward to find thanks to the analytical model we established by assuming that the net thermal energy extracted from the thermal reservoirs is fully converted into the coherent motion of the driving mode. By deriving a dynamical non-Markovian quasiclassical model and utilizing it in demonstrating the operation of the quantum heat engine, we justified the assumption made in the analytical model. We achieved output powers and efficiencies of the heat engine in a very good agreement between the two models. The efficiency obtained was roughly 10\% and the output power was found to be a reasonable fraction of the energy quantum of the working fluid times the frequency of the driving mode. Importantly, the output power was found to monotonically increase with the temperature of the hot reservoir providing prospects for possible applications.

Our initial analysis showed that the output power of the quantum heat engine significantly fluctuates from noise realization to another. Although this fluctuation seems suppressed in comparison to Poisson processes, it calls for future studies on the noise spectrum of the coherent-field output of the heat engine, and possible ways to decrease the observed phase noise and power fluctuations.  

Thanks to the versatility of the proposal, we consider a future experimental realization of our quantum-heat-engine proposal likely feasible, but requiring prior specific theoretical analysis for each experimental realization. Although we leave such specialized analysis for future work, we note that our example calculations show that the quality factor the the working-fluid mode may be of the order of hundred and that of the driving mode roughly thousand for the optomechanical coupling constant of roughly one percent of the angular frequency of the working fluid. In typical optomechanical systems, for example, such quality factors are straightforward to achieve, but the optomechanical coupling constant may be smaller, which calls for an optimization of the separation of the reservoir spectra~\cite{optomech, qoptom, Meystre}. In addition to optomechanical systems, it is interesting to consider a realization of the quantum heat engine in superconducting circuits~\cite{Pekola2015, Thomas, Pekola, Hardal2017} which may offer more flexibility in engineering the optomechanical coupling constant~\cite{Johansson, qoptom, Ranni} and angular frequencies of the two modes~\cite{Chen, Ranni}. For high-power operation, it may be interesting to consider engineered environments~\cite{Tan2017, Timm_review} which have recently been demonstrated to provide fast preparation of thermal states~\cite{Timm_thermal}.

Even with the possible loss of quantum character due to our quasiclassical methods of solution, we deemed retaining the non-linearity and non-Markovianity as the important novelty of our approach, and although investigating the full quantum character of the device is left for future work, the results obtained here remain significant and interesting proof of concept. In the future, it is interesting to study several different possible experimental realizations of the proposed quantum heat engine and their advantages and disadvantages with respect to each other. In addition, the mitigation of the phase and amplitude noise in the coherent output field of the heat engine call for further studies together with possible use cases of this work obtained from the quantum heat engine.

\section*{Acknowledgements}

This work was funded by the Academy of Finland Centre of Excellence program (project Nos. 352925, and 336810) and grant Nos.~316619 and 349594 (THEPOW). We also acknowledge funding from the European Research Council under Advanced Grant No.~101053801 (ConceptQ). 

We thank Vasilii Vadimov, Jukka Pekola, and Bayan Karimi for scientific discourse.

\bibliographystyle{unsrt}
\bibliography{refs}
\clearpage

\section*{Supplementary Material}


This Supplementary Materials contains the following figures that provide additional complementary data for the main text, but that are not pivotal for the conclusions of the work: temporal evolution of the optomechanical system free of thermal fluctuation (Fig.~S(1)) dynamics of the optical and mechanical mode amplitudes in 2+1 dimensions (Fig.~S(2)), dependence on the output power in the analytical model on the dissipation rates (Fig.~S(3)), the dependence of the energy produced in a cycle on the mechanical-mode frequency (Fig.~S(4)), and the dependence of the interaction time between the working fluid and its reservoirs on the speed of the driving mode (Fig.~S(5)). Figure~S(6) shows how the interaction time $\tau$ is defined for the analytical model. 

\renewcommand{\figurename}{Supplementary Figure S(\arabic{figure})}
\renewcommand{\thefigure}{\!}
\setcounter{figure}{0}

\begin{figure}[!ht]
\centering
\includegraphics[width=0.6\textwidth, trim={0 0 0 0}, clip]{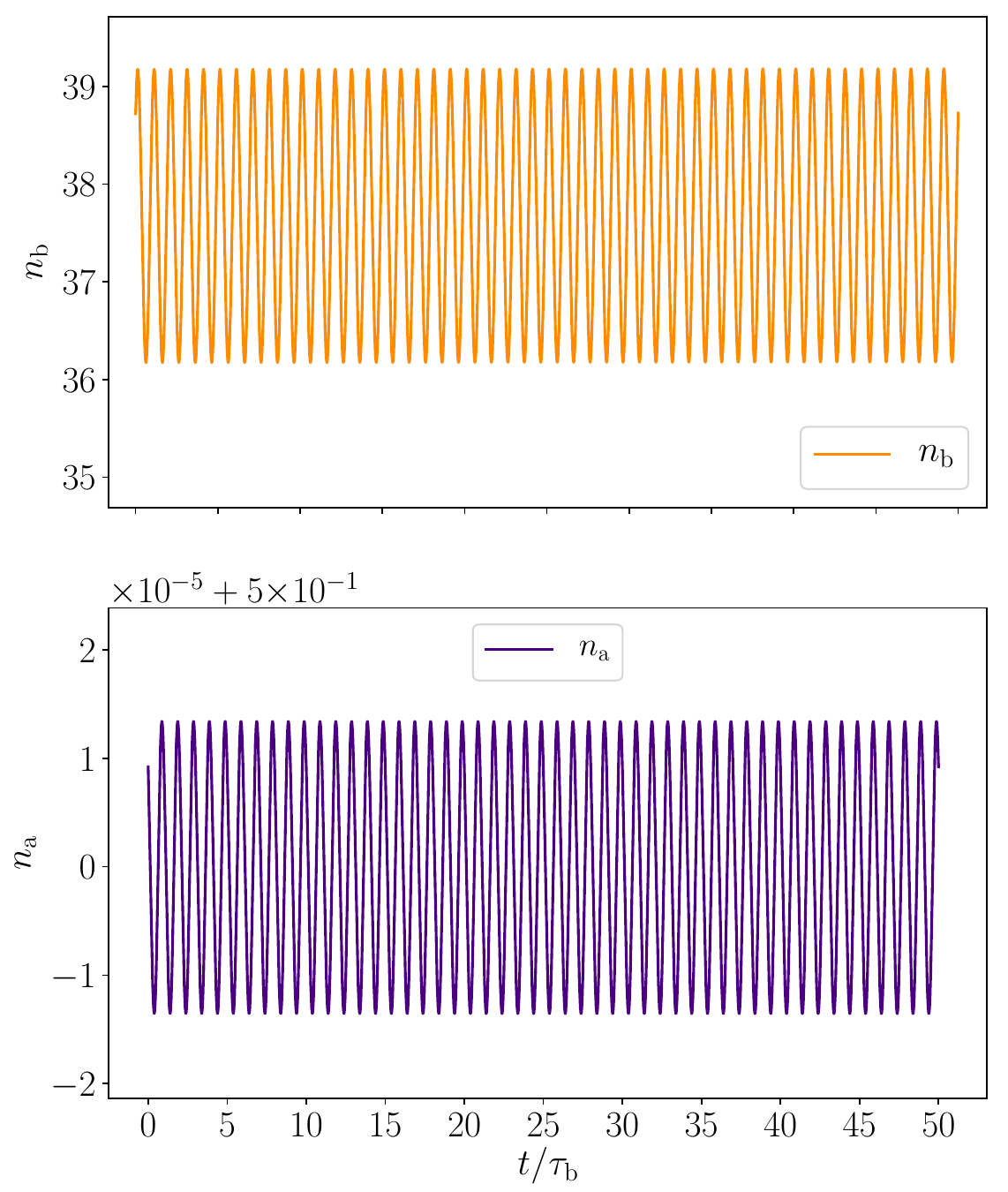}
\caption{\textbf{Temporal evolution of the optomechanical system free of thermal fluctuation.} Temporal evolution of the mean occupation numbers of the optomechanical system modes, orange for the mechanical mode and magenta for the optical, uncoupled from the thermal noise sources. We use identical parameters values to those in Fig.~\ref{time_evol}.}
\label{fig:s6}
\end{figure}

\begin{figure}[!ht]
\centering
\includegraphics[width=0.7\textwidth, trim={0 0 0 0}, clip]{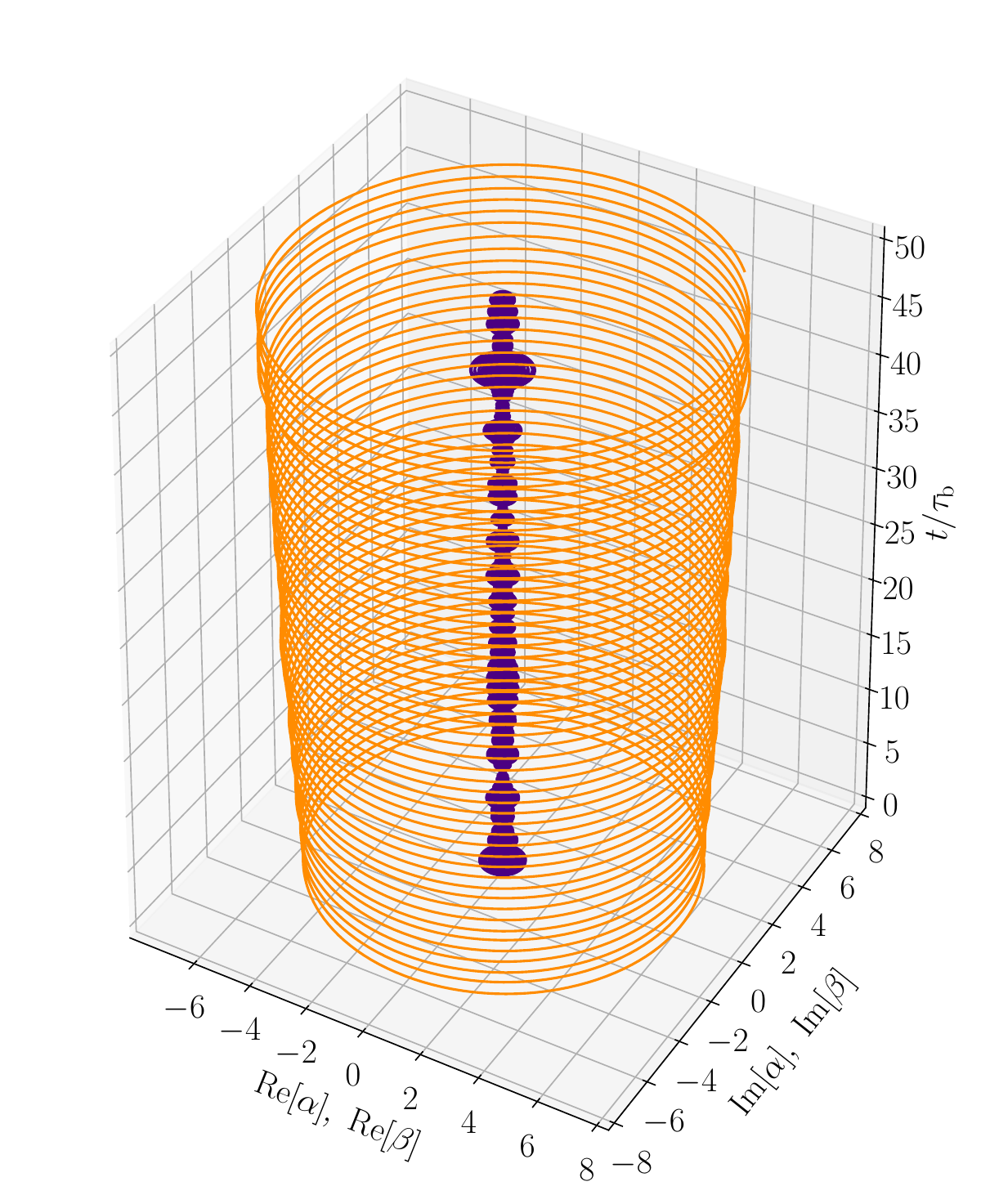}
\caption{\textbf{Dynamics of the amplitudes of the optical and mechanical modes.} Phase space evolution of the complex-valued amplitudes of the optical (purple color) and the mechanical (orange color) mode illustrated in two plus one dimension. These data is identical to those shown in Figs.~\ref{time_evol}(\textbf{b}) and~\ref{time_evol}(\textbf{d}).}
\label{3d}
\end{figure}

\begin{figure}[!ht]
\centering
\includegraphics[width=0.7\textwidth, trim={0 0 0 0}, clip]{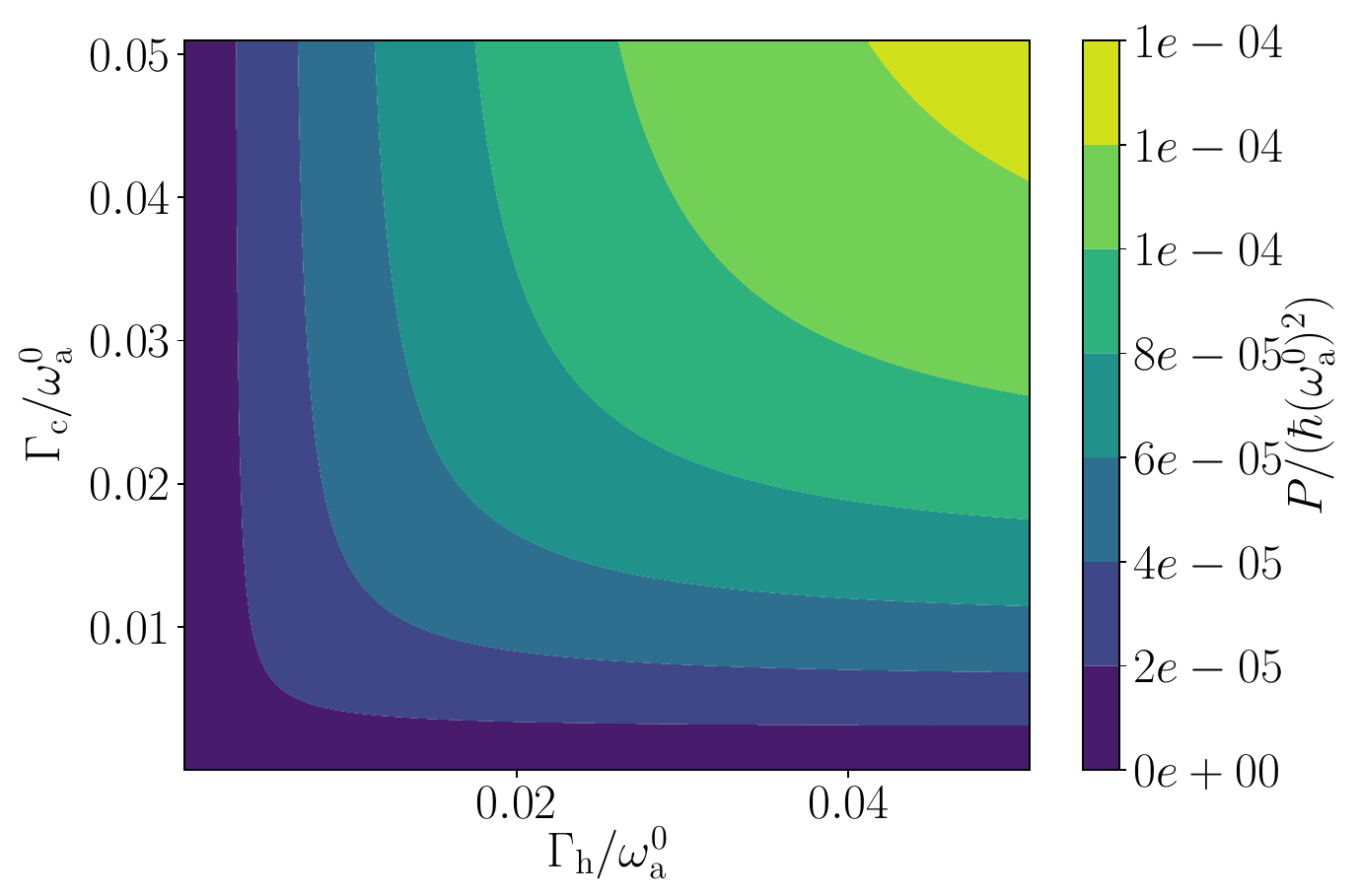}
\caption{\textbf{Effect of the reservoir coupling strengths on the output power.} Net output power of the quantum heat engine as a function of the energy relaxation rates $\Gamma_{\uh/\uc}$ given by the analytical model. We use identical parameters values to those in Fig.~\ref{fig:heuristic_params}.}
\label{fig:s2}
\end{figure}

\begin{figure}[!ht]
\centering
\includegraphics[width=0.7\textwidth, trim={0 0 0 0}, clip]{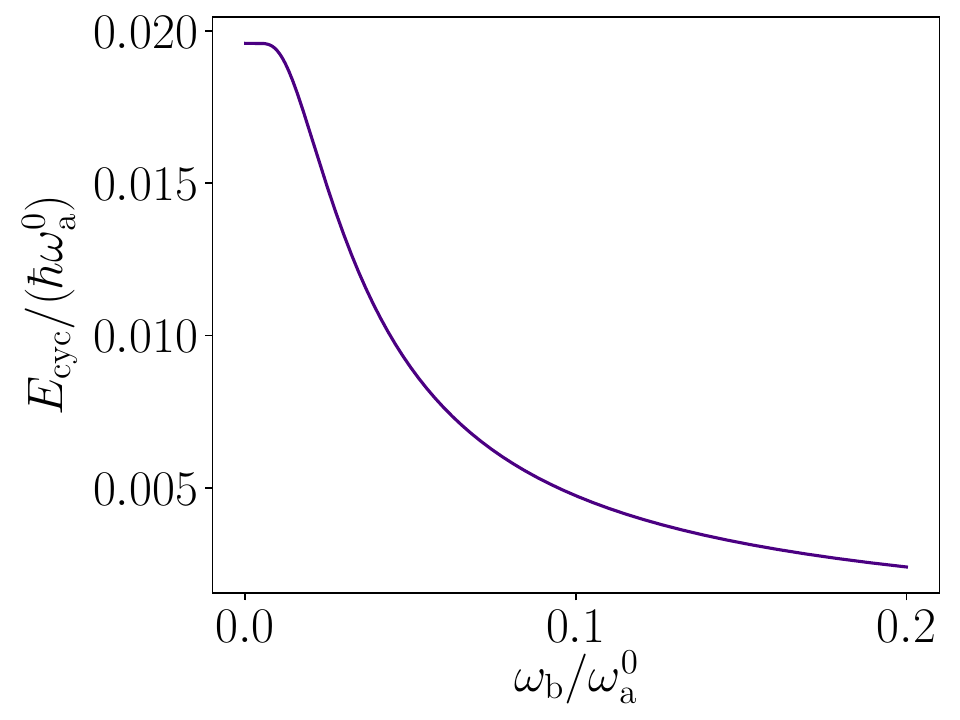}
\caption{\textbf{Energy produced in a cycle.} Energy per cycle as given by the analytical model Eq.~\eqref{ecyc} as a function of mechanical-mode angular frequency $\omega_\ub$. We use identical parameters values to those in Fig.~\ref{fig:heuristic_params}.}
\label{fig:s3}
\end{figure}

\begin{figure}[!ht]
\centering
\includegraphics[width=0.7\textwidth, trim={0 0 0 0}, clip]{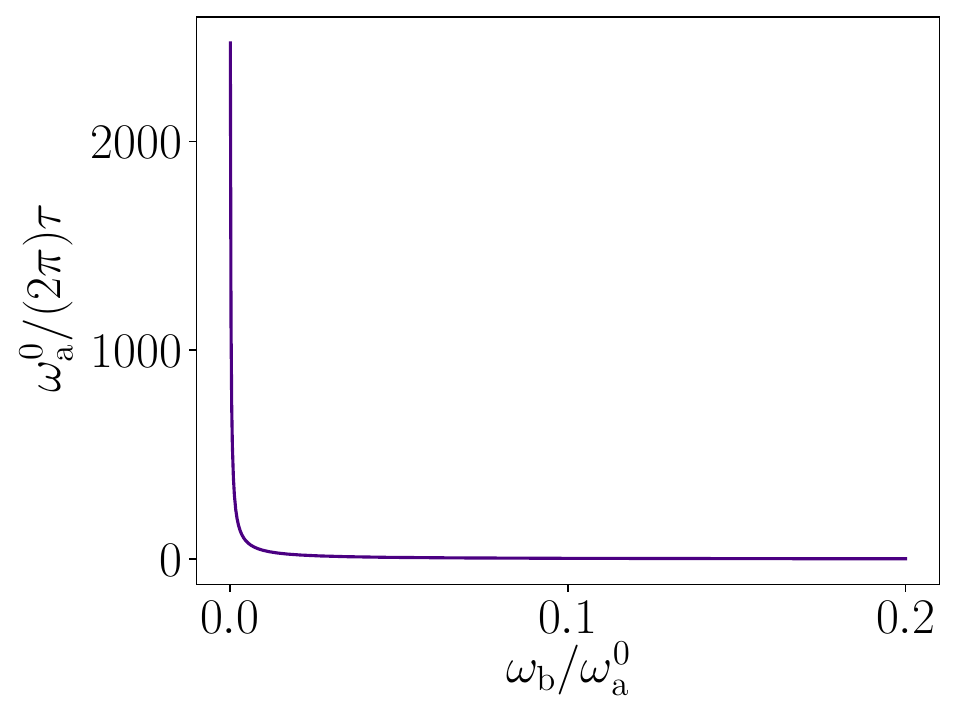}
\caption{\textbf{Dependence of the interaction time between the working fluid and its reservoirs on the speed of the driving mode.} Interaction time $\tau$ of the analytical model as a function of the mechanical-mode angular frequency $\omega_\ub$. We use identical parameters values to those in Fig.~\ref{fig:heuristic_params}.}
\label{fig:s4}
\end{figure}

\begin{figure}[!ht]
\centering
\includegraphics[width=0.7\textwidth, trim={0 0 0 0}, clip]{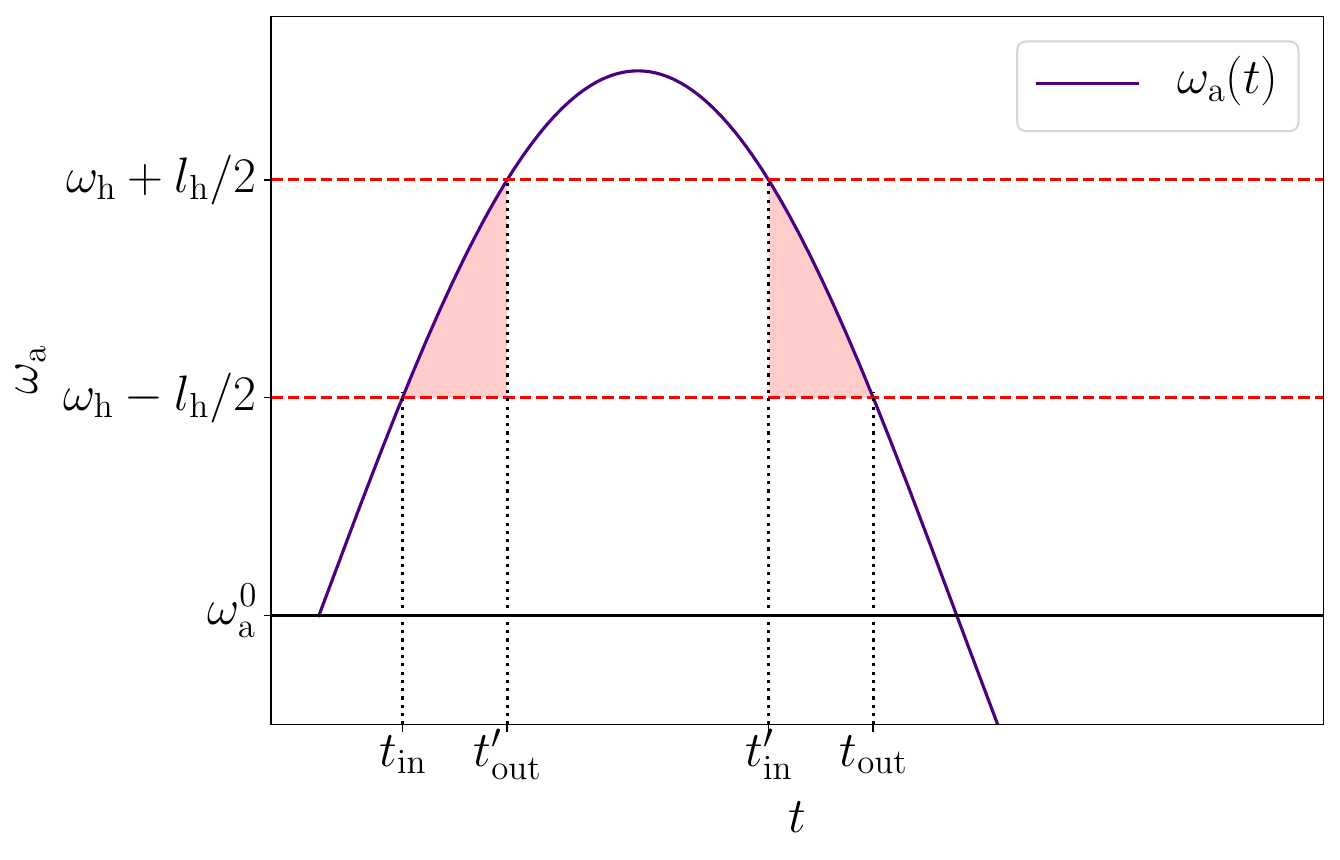}
\caption{\textbf{Calculation of the interaction time $\tau$ of the analytical model.} Optical-mode angular frequency $\omega_\mathrm{a}(t)$ as a function time (purple solid line). The shaded areas represent the interaction regions of the optical mode with the hot reservoir, defining the interaction time $\tau=(t_\uout-t_\uin)-(t_\uin'-t_\uout')$, where the existence of each term in parenthesis depends on the amplitude of the angular-frequency modulation.}
\label{fig:s5}
\end{figure}

\clearpage

The interaction time $\tau$ of the analytical model for the hot reservoir in the most general case can be calculated, as depicted in Fig.~S(5), from the inequality
\begin{align*}
\omega_\uh-l_\uh/2\leq \omega_\ua^0+\Delta\omega_\ua/2\sin(\omega_\ub t)\leq \omega_\uh+l_\uh/2.
\end{align*}
Solving the two equations arising at the limit of equality, we find the conditions
\begin{align*}
t_\uin&=\frac{1}{\omega_\ub}\arcsin(\frac{2\omega_\uh-l_\uh-2\omega_a^0}{\Delta\omega_\ua}),\\
t_\uout&=\frac{1}{\omega_\ub}\left[\pi-\arcsin(\frac{2\omega_\uh-l_\uh-2\omega_a^0}{\Delta\omega_\ua})\right],\\
t_\uout'&=\frac{1}{\omega_\ub}\arcsin(\frac{2\omega_\uh+l_\uh-2\omega_a^0}{\Delta\omega_\ua}),\\
t_\uin'&=\frac{1}{\omega_\ub}\left[\pi-\arcsin(\frac{2\omega_\uh+l_\uh-2\omega_a^0}{\Delta\omega_\ua})\right].
\end{align*}
Using these definitions, the interaction time can be calculated from $\tau=(t_\uout-t_\uin)-(t_\uin'-t_\uout')$. 

The average angular frequency over the interaction period $\bar{\omega}_\ua^\uh$ in the analytical model is found by time averaging over the interaction period,
\begin{align*}
\bar{\omega}_\ua^\uh=\frac{1}{\tau}\int_{t_\uin}^{t_\uout}\left[\omega_\ua^0+\frac{\Delta\omega_\ua}{2}\sin(\omega_\ub t)\right]\left[\Theta(t_\uout'-t)+\Theta(t-t_\uin')\right]\ud t,
\end{align*}
where $\Theta(t)$ is the Heaviside step function. Analogous calculations apply for the cold reservoir.

\end{document}